\documentclass[prl,twocolumn,nopacs,nofootinbib,superscriptaddress]{revtex4-2}

\usepackage{graphicx}
\usepackage{dcolumn}

\usepackage{color}

\usepackage[dvips]{epsfig}
\usepackage{amssymb}

\usepackage{amsmath}

\usepackage{eucal}
\usepackage{mathrsfs}
\usepackage{amsmath}
\usepackage{amssymb}
\usepackage{amsfonts}
\usepackage{latexsym}
\usepackage{color}

\newcommand{\be}{\begin{eqnarray}}
\newcommand{\ee}{\end{eqnarray}}
\newcommand{\bse}{\begin{subequations}}
\newcommand{\ese}{\end{subequations}}

\newcommand{\bnum}{\begin{enumerate}}
\newcommand{\enum}{\end{enumerate}}

\newcommand{\bit}{\begin{itemize}}
\newcommand{\eit}{\end{itemize}}

\newcommand{\bc}{\begin{cases}}
\newcommand{\ec}{\end{cases}}

\newcommand{\bpm}{\begin{pmatrix}}
\newcommand{\epm}{\end{pmatrix}}

\newcommand{\bvm}{\begin{vmatrix}}
\newcommand{\evm}{\end{vmatrix}}

\newcommand{\ran}{\rangle}

\begin{document}

\title{Harmonic flow field representations of quantum bits and gates}

\author{Vishal P. Patil}
\address{Department of Mathematics, 
Massachusetts Institute of Technology, 
77 Massachusetts Avenue, 
Cambridge, MA 02139}
\address{School of Humanities and Sciences, 
Stanford University,
450 Serra Mall, Stanford, CA 94305}
\author{\v{Z}iga Kos}
\address{Department of Mathematics, 
Massachusetts Institute of Technology,  
77 Massachusetts Avenue, 
Cambridge, MA 02139}
\address{Faculty of Mathematics and Physics,
University of Ljubljana,
Jadranska 19, 1000 Ljubljana, Slovenia}
\author{J\"orn Dunkel}
\email{dunkel@mit.edu}
\address{Department of Mathematics, 
Massachusetts Institute of Technology, 
77 Massachusetts Avenue, 
Cambridge, MA 02139}
\date{\today}

\begin{abstract}
We describe a general procedure for mapping arbitrary $n$-qubit states to two-dimensional (2D) vector fields. The mappings use complex rational function representations of individual qubits, producing classical vector field configurations that can be interpreted in terms of  2D inviscid fluid flows or electric fields. Elementary qubits are identified with localized defects in 2D  harmonic vector fields, and multi-qubit states find natural field representations  via  complex superpositions of vector field products. In particular, separable states appear as highly symmetric flow configurations, making them both dynamically and visually distinct from entangled states. The resulting real-space representations of entangled qubit states enable an intuitive visualization of their transformations under quantum logic operations. We demonstrate this for the quantum Fourier transform and the period finding process underlying Shor's algorithm, along with other quantum algorithms. Due to its generic construction, the mapping procedure suggests the possibility of extending concepts such as entanglement or entanglement entropy to classical continuum systems, and thus may help guide new experimental approaches to information storage and non-standard computation.
\end{abstract}


\maketitle

\newcommand{\todo}[1]{{\color{blue} ToDo: #1}}
\newcommand{\jd}[1]{{\color{red}#1}}


Classical continuum systems can be used to store and manipulate discrete information robustly. From  passive~\cite{prakash2007microfluidic,gilpin2018cryptographic,ChiuDT_ProcNatlAcadSci98_2001} and active~\cite{gunji2011robust,woodhouse2017active,NicolauDV_ProcNatlAcadSci113_2016} fluids to DNA-based computers~\cite{qian2011scaling}, the large configuration spaces of continuum systems allow for the stable creation and handling of classical bits~\cite{fruchart2020dualities}. A particularly interesting question is how this information storage capability compares to quantum systems~\cite{Ladd:2010aa} which typically involve entanglement~\cite{nematic_bits}. Recent work has shown that certain non-quantum entanglements~\cite{karimi2015classical} can be realized in classical light fields~\cite{spreeuw1998classical,goldin2010simulating,song2015bell,kedia2013tying} and coupled billiard balls~\cite{BrownAR_Quantum4_2020}. However, the problem of creating analogues of more general entanglements in classical continuum systems continues to present challenges~\cite{nematic_bits,BushJWM_RepProgPhys84_2020,BushJWM_Chaos28_2018}. Here, by building on ideas underlying the Poincar\'e-Bloch sphere~\cite{makela2010n,makela2010polynomial} and Majorana star~\cite{bruno2012quantum,kam2020three,devi2012majorana,serrano2020majorana} representations of quantum states, we exhibit a family of formal mappings between $n$-qubit states and 2D hydrodynamic flow fields. The topological defects and stagnation points of these flow fields encode properties of the corresponding $n$-qubit state.

\begin{figure}[b]
\centering
\includegraphics[width=\columnwidth]{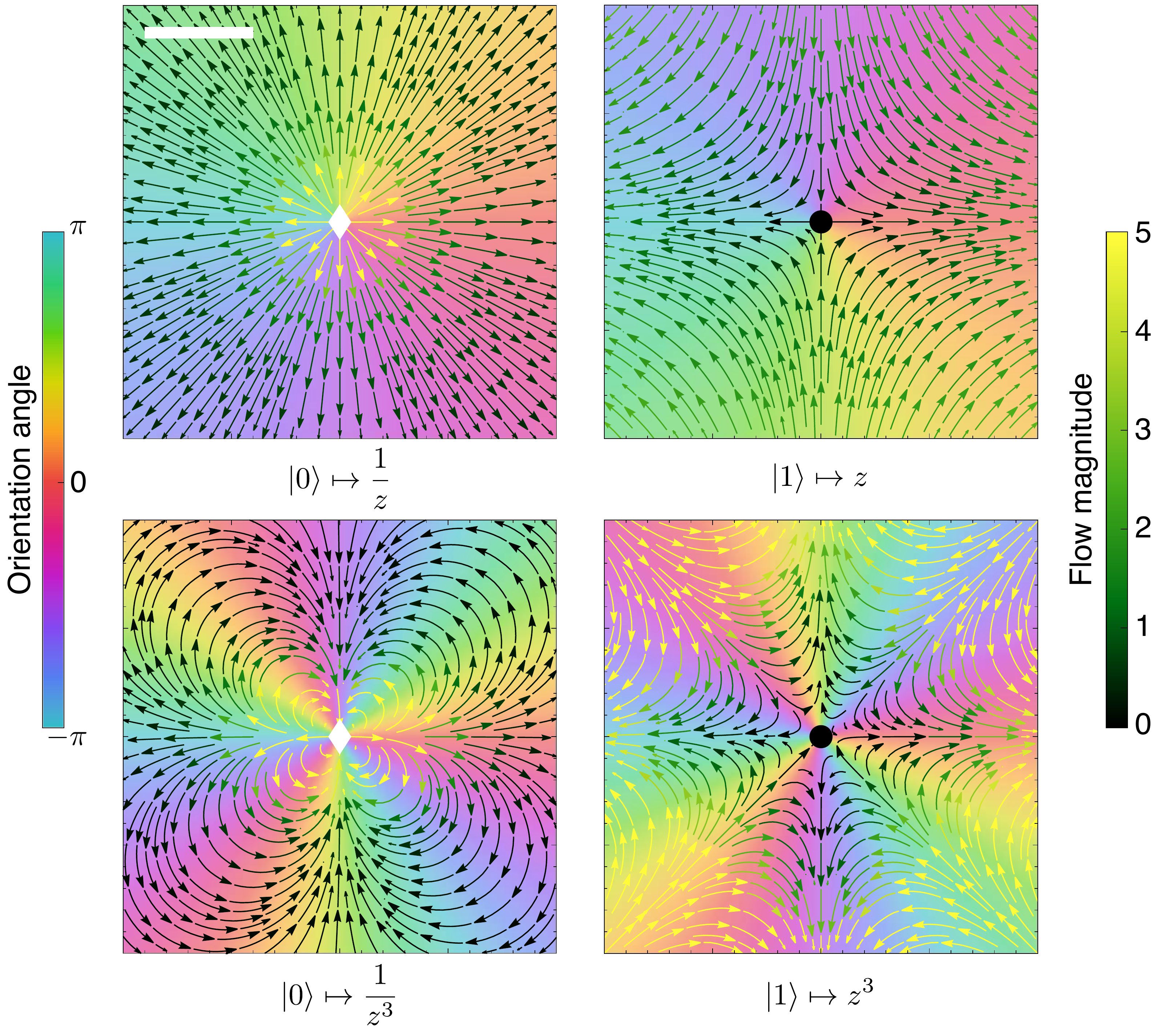}
\caption{Vector field representations of single-qubit basis states. The basis states are mapped to complex functions $f$ which correspond to 2D vector fields $(u,v) = (\text{Re}\,f^*, \text{Im}\,f^*)$. In complex notation the flow fields have a single zero (black circle) or a pole (white diamond). Poles indicate the state $|0\rangle \sim z^{-d}$, and zeros indicate the state $|1\rangle\sim z^{d}$, illustrated here for defect charge $d=1$ and $d=3$. Scale bar: length~1.} 
\label{fig1}
\end{figure}

\begin{figure*}[t]
	\centering
	\includegraphics[width=\textwidth]{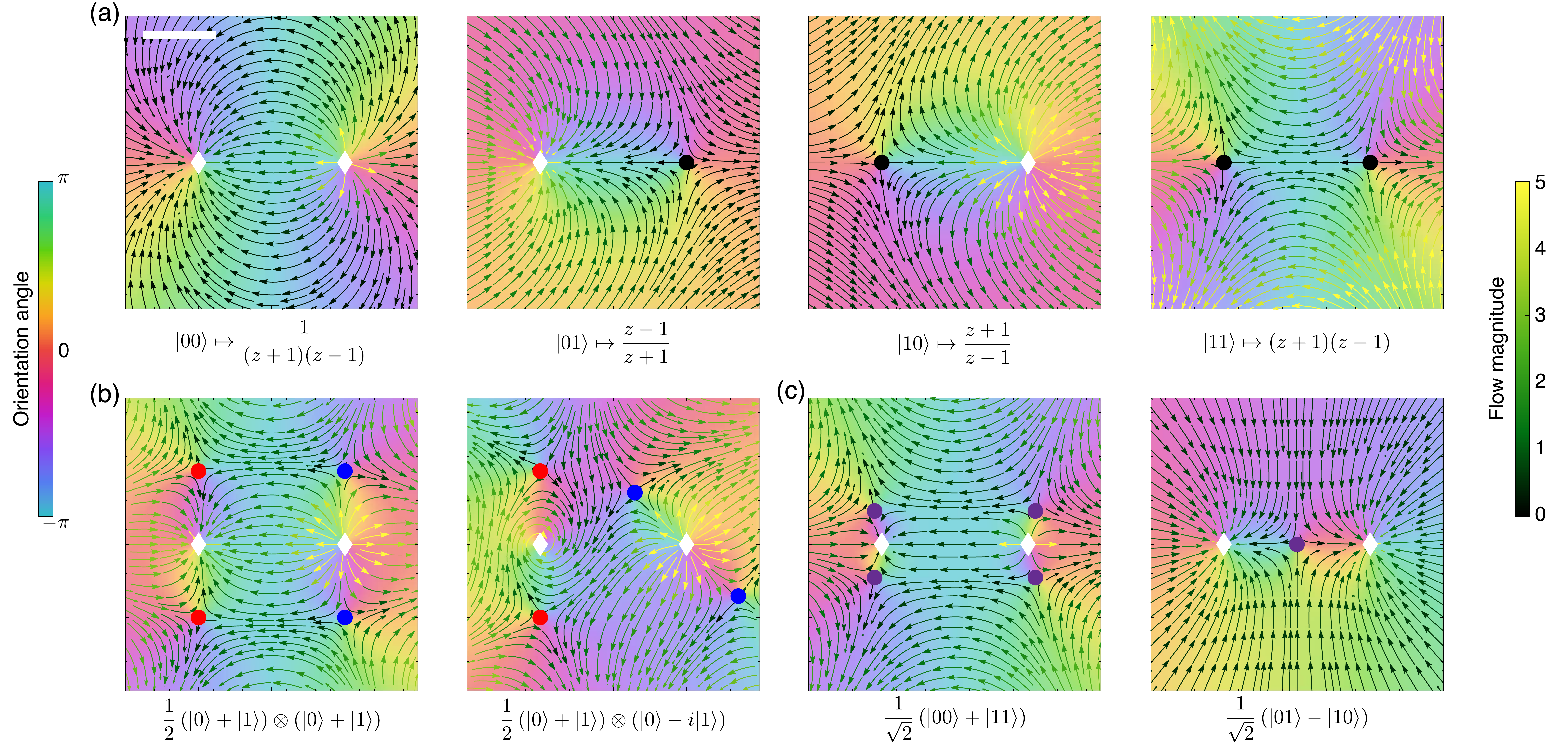}
	\caption{Vector field representations of 2-qubit states.
		(a)~The 2-qubit computational basis states are mapped to complex functions $f$, with poles (white diamonds) and zeros (filled circles) at the locations of the basis defects (here, $-1$ and $+1$). The basis states have defects with fixed absolute charge $d$ (here $d=1$). 
		(b)~Separable states typically give rise to poles at the basis defect positions (diamonds) surrounded by equiangular halos of stagnation points (filled, color-coordinated circles) where the flow velocity vanishes. The number of halo points is equal to $2d$. The separable states in (a) arise when the halo defects are at infinity or coincide with the locations of the basis defects. 
		(c)~Entangled non-separable states, such as the Bell states depicted here, are distinguished by their lack of regular defect halos. Scale bar: length~1.} 
	\label{2_qubit_fig}
\end{figure*}
\par
The resulting classical qubit representations have the property that the tensor product of qubits corresponds to the complex product of vector fields, and qubit superpositions correspond to sums of vector fields. Furthermore, by using analytic functions with a finite number of singular points as the elementary building blocks, these mappings represent multi-qubit states as harmonic vector fields, which can theoretically be realized as inviscid potential flows. In the mapping presented here, qubit properties are linked to the charge and position of defects (although, in principle, infinitely many other representations could be constructed in a similar manner [SI]). Entanglement corresponds to vector fields with irregular patterns of zeros, whereas separable states consist of defects arranged in regular polygonal halos around central basis defects. Thus, these results additionally present an alternative perspective on the role of topological defects in 2D flows~\cite{crowdy2010new}. Below, we demonstrate how this representation can be used to visualize the quantum Fourier transform~\cite{CleveR_Proceedings454_1998} and the period funding subroutine of Shor's algorithm~\cite{shor1999polynomial}. The complexity of the underlying fluid structures illustrates the difficulty of simulating a quantum computer with a classical device.

\par

Our hydrodynamic visualizations of multi-qubit states are based on mappings from a Hilbert space to the vector space of 2D vector fields $ (u(x,y), v(x,y))$. Since points $(x,y) \in \mathbb{R}^2$ and vectors $(u,v) \in \mathbb{R}^2$ can be identified with complex numbers, this space, $\mathcal{V}$, is also the vector space of complex functions
\begin{align*}
z = x+iy &\mapsto f(z) = u(x,y) - i v(x,y)
\end{align*}
Restricting to holomorphic functions $f$ selects for harmonic vector fields $(\text{Re}\,f^*, \text{Im}\,f^*)=(u,v)$, which are both curl-free and divergence-free away from singularities. This fact follows from the Cauchy-Riemann equations for~$f$, which imply $u_x+v_y = u_y-v_x = 0$. Such vector fields  arise in a variety of classical 2D phenomena, including fluid flows~\cite{crowdy2010new}, electrostatic fields~\cite{furman1994compact}, polar liquid crystals~\cite{khoromskaia2017vortex,TangX_SoftMatter13_2017} and elastic systems~\cite{sarkar2021elastic,muskhelishvili1963some}. The zeros and poles of $f$ are defects of the vector field~(Fig.~\ref{fig1}), and facilitate visualizations of the qubit states.

\par

To visualize qubits as 2D vector fields, we construct a linear map, $\Phi^{(n)}$, from the space of $n$ qubits, $\mathcal{H}^{(n)}$, to the vector space of complex functions which are smooth almost everywhere, $\mathcal{V}$, in such a way that the tensor product of qubits corresponds to a product of complex functions. Since the tensor product of qubits is not commutative, we must construct a sequence of linear maps $\phi_j$ for $1\leq j\leq n$ which send the 1-qubit Hilbert space $\mathcal{H}=\mathcal{H}^{(1)}$ to different subspaces of complex functions, $\phi_j(\mathcal{H}) \subset \mathcal{V}$. In particular, $\phi_j(\mathcal{H}) \neq \phi_k(\mathcal{H})$ if $j\neq k$. From the base maps $\phi_j:\mathcal{H}\rightarrow \mathcal{V}$, the linear map $\Phi^{(n)}:\mathcal{H}^{(n)} \rightarrow\mathcal{V}$ can be constructed as
\begin{align*}
\sum_{\boldsymbol{\sigma}\in\{0,1\}^n} \lambda_{\boldsymbol{\sigma}}|\sigma_1\sigma_2...\sigma_n\rangle &\mapsto \sum_{\boldsymbol{\sigma}\in\{0,1\}^n} \lambda_{\boldsymbol{\sigma}} \prod_{j=1}^n \phi_j\big(|\sigma_j\rangle\big)
\end{align*}
where $\lambda_{\boldsymbol{\sigma}} \in \mathbb{C}$ is the component of the computational basis state $|\boldsymbol{\sigma}\rangle = |\sigma_1\sigma_2...\sigma_n\rangle$. The linearity of $\Phi^{(n)}$ ensures that superposition of qubits corresponds to addition of vector fields. For mathematical consistency, the linear independence of different qubits must be preserved in the vector field representation. This condition sets $\ker\left(\Phi^{(n)}\right) = \{\mathbf{0}\}$. The computational basis states are built up by taking products of the functions $\phi_j\big(|\sigma\rangle\big)$.
In other words, the tensor product operation becomes complex multiplication in the vector field representation.

\begin{figure}[t]
	\centering
	\includegraphics[width=\columnwidth]{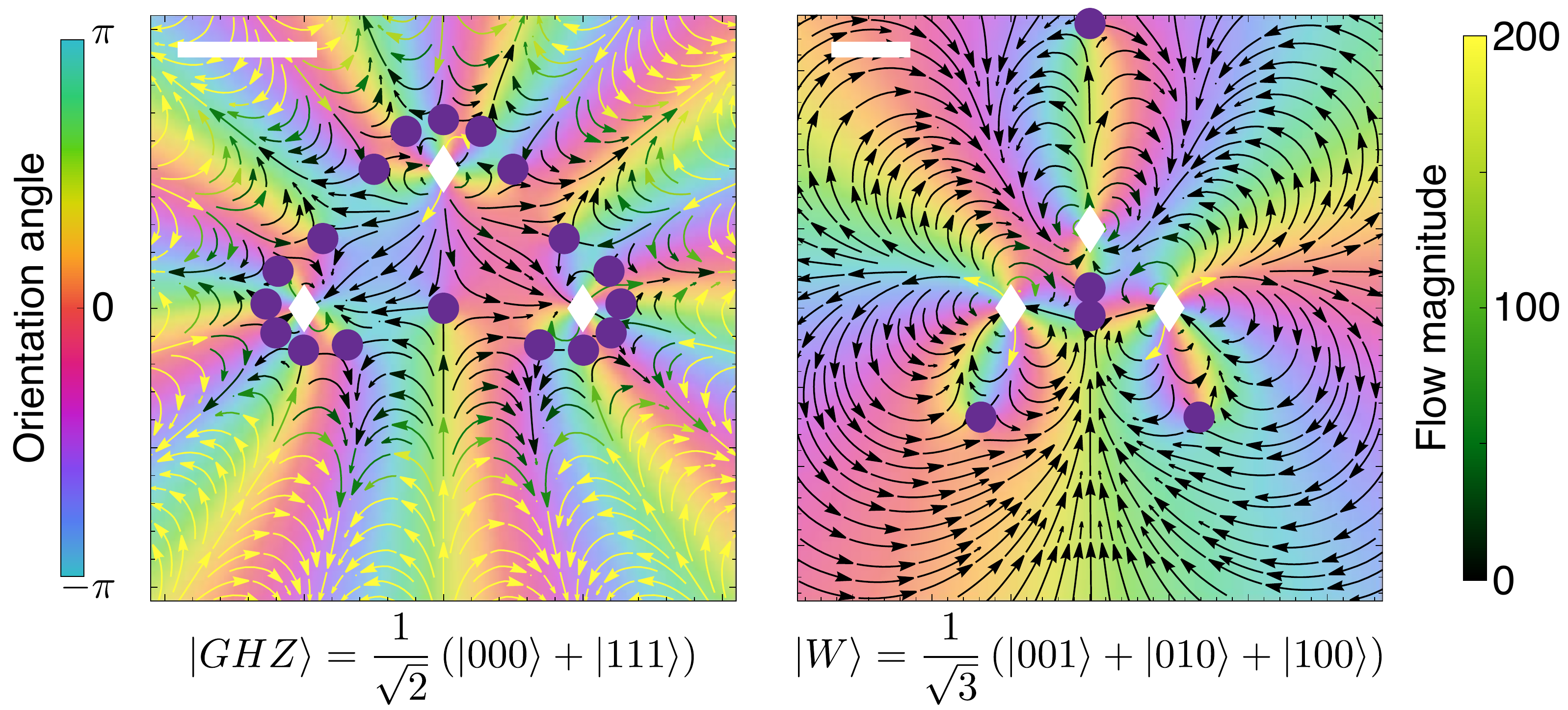}
	\caption{3-qubit entanglement. Entangled 3-qubit states correspond to flows with irregular patterns of zeros (purple circles) around the basis defects at $-1,i,+1$. Visualization has charge $d=3$. White diamonds denote poles. Scale bars: length~1.} 
	\label{3_qubit_fig}
\end{figure}

\begin{figure}[t]
	\centering
	\includegraphics[width=\columnwidth]{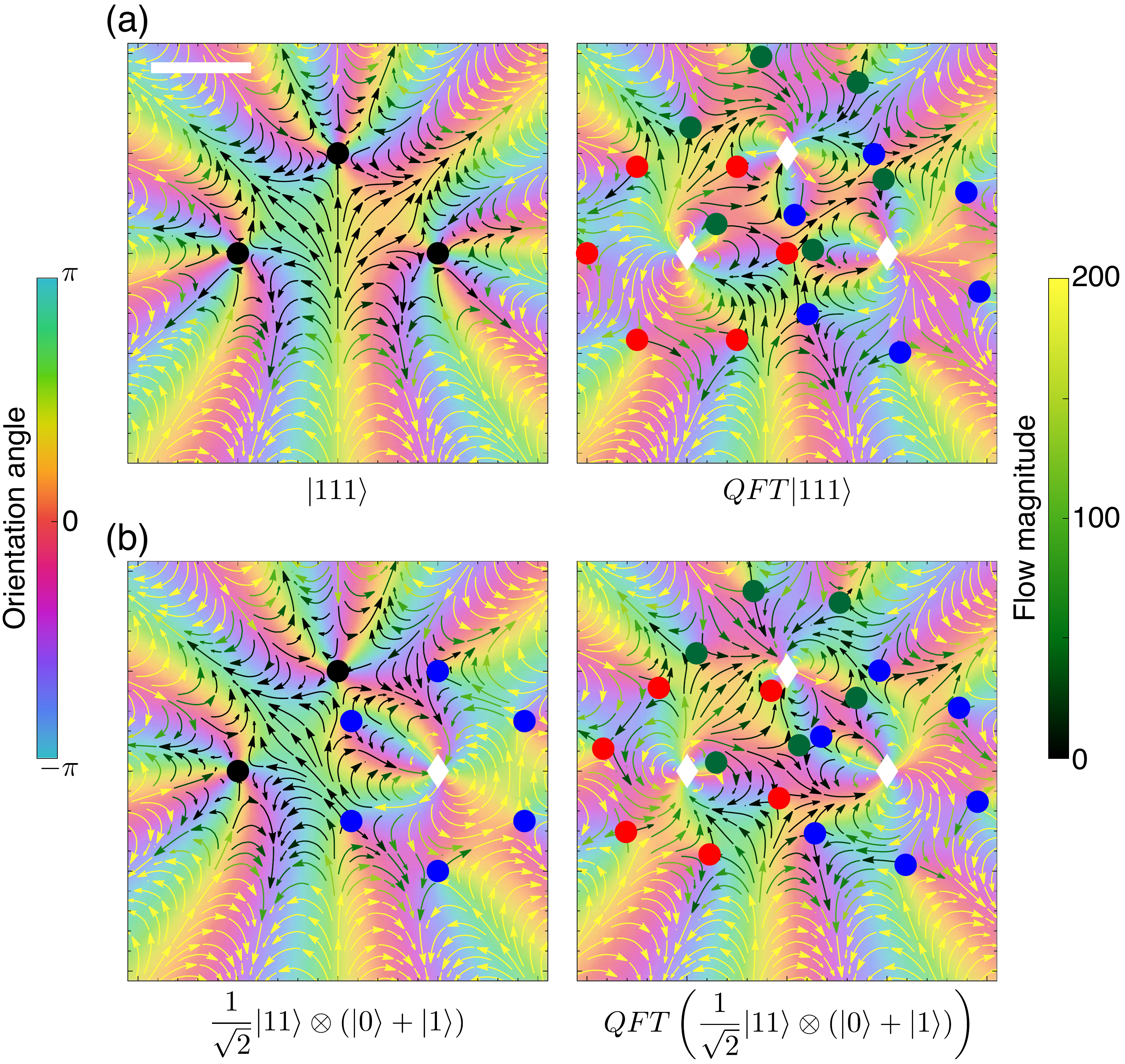}
	\caption{Quantum Fourier transform. (a)~The quantum Fourier transform~\cite{CleveR_Proceedings454_1998} acts on 3 qubits at $-1,+i,+1$ with defect charge $d=3$. Computational basis states are mapped to separable states with equiangular defect halos of size $2d=6$ (red, green and blue circles), centered at the locations of the basis defects.
	(b)~The quantum Fourier transform of non-basis states generates entanglement, which is apparent from the irregular, non-equiangular pattern of defects in the flow field picture. Halo defects are colored red, green or blue by interpolating from $QFT|111\rangle$ (top row) to $QFT\left(|111\rangle + |110\rangle\right)$ (bottom row) in the flow field picture. Scale bar: length~1.} 
	\label{qft_fig}
\end{figure}

Using this class of harmonic vector field representations, we can realize qubit indistinguishability by selecting maps which send different qubits to translations of the same elementary basis vector field~ (Fig.~\ref{2_qubit_fig}). A general construction of such a representation starts with choosing two functions $q_0(z)$ and $q_1(z)$ to represent the $|0\rangle $ and $|1\rangle$ states,  respectively. An $n$-qubit state in the computational basis is then constructed by taking translations and products of $q_0$ and $q_1$
\begin{align*}
\phi_j\big(|\sigma\rangle\big) = q_{\sigma}(z - a_j), \qquad |\sigma_1\sigma_2...\sigma_n\rangle = \prod_{j=1}^n q_{\sigma_j}(z-a_j)
\end{align*}
where the $a_j$'s are distinct but can otherwise be chosen arbitrarily. From this basis, any $n$-qubit state may be obtained by addition. In particular, we focus on the case where $q_0(z) = z^{-d}$ and $q_1(z) = z^d$ (Fig.~\ref{2_qubit_fig}a), which corresponds to storing qubit information in the degree and location of a topological defect. The $a_j$'s are the locations of the basis defects of the flow field, and we refer to $d$ as the charge of the visualization. Since the state $|0\rangle$ is mapped to $(z-a_j)^{-d}$, superpositions will typically have poles at the basis defect locations (Figs.~\ref{2_qubit_fig}b and \ref{2_qubit_fig}c). For linear independence, the charge $d$ must depend on $n$, the number of qubits which are being represented [SI]. In general, $d$ grows exponentially with $n$, but for $n\leq 4$, it suffices to choose $d\leq 3$ (Figs.~\ref{3_qubit_fig} and \ref{qft_fig}). Specifically, for large $n$, we find that a necessary condition for linear independence is $d\sim n^{-1} 2^n$ [SI]. On the other hand, a representation which stored all qubit properties in the defect degree would require $d \geq 2^n$ [SI]. This further illustrates the practical limitations of experimentally implementing large $n$-qubit systems using  classical continuum fields. Despite these caveats, the fluid flow representation appears experimentally viable for smaller values of $n$. For example, the commonly used entangled 3-qubit $|GHZ\rangle$ and $|W\ran$ states can be realized with three moderately charged defects (Fig.~\ref{3_qubit_fig}).

An immediate benefit of the vector field representation is that entanglement can be detected through direct visual inspection of flow patterns. Since tensor products correspond to vector field products, entanglement is linked to an asymmetry in the halos of stagnation points that surround the basis defects. To see this,  consider the space of 2-qubit vector fields $\Phi^{(2)}\left(\mathcal{H}^{(2)}\right)$ with charge $d=1$ and basis defects at $\pm 1$ (Fig.~\ref{2_qubit_fig}) given by 
\begin{align*}
    |\sigma_1 \sigma_2\rangle \mapsto (z+1)^{2\sigma_1 -1} (z-1)^{2\sigma_2 -1}
\end{align*}
A general separable state in $\mathcal{H}^{(2)}$ corresponds to a rational function as follows
\begin{align*}
    \Phi^{(2)}\Big( \left( \alpha |0\rangle + \beta |1\rangle \right) \otimes \left(\gamma |0\rangle + \delta |1\rangle \right) \Big) = f(z)
\end{align*}
where
\begin{align*}
    f(z) = \frac{\left( \alpha + \beta(z+1)^2 \right) \left( \gamma + \delta (z-1)^2\right)}{(z+1)(z-1)}
\end{align*}
This yields a vector field with halo defects at $-1\pm \sqrt{-\alpha / \beta}$ and $1\pm \sqrt{-\gamma/\delta}$, and these defects are zeros of the corresponding holomorphic function. In special cases, these halos of zeros around $\pm 1$ could be pushed to infinity or could coincide with either of the basis defects at $\pm 1$. For example, the separable basis states $|00\rangle$ and $|11\rangle $ have $\beta=\delta=0$ and $\alpha=\gamma=0$ respectively, which correspond to halo defects at infinity, or at $\pm 1$. However, a generic separable state will have an equiangular halo of defects arranged around the basis defect locations (Fig.~\ref{2_qubit_fig}b), with no additional free defects [SI]. Entangled states, which lack these regular halo defects, are therefore characterized by their asymmetric defect patterns (Figs.~\ref{2_qubit_fig}c and~\ref{3_qubit_fig}).

\begin{figure*}[t]
	\centering
	\includegraphics[width=\textwidth]{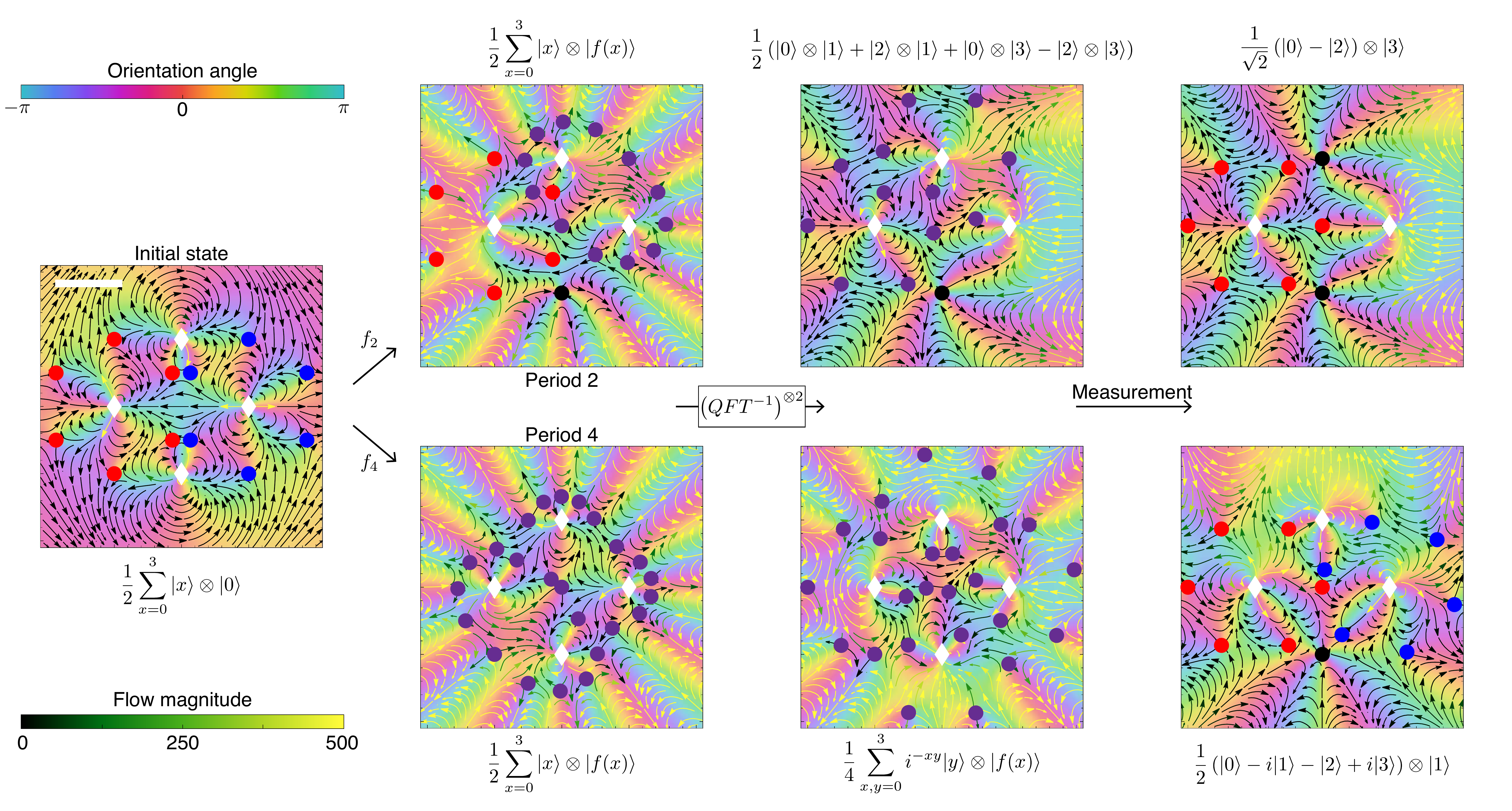}
	\caption{Flow field visualization of the period finding subroutine in Shor's algorithm.
		The quantum period finding algorithm on 4 qubits with charge $d=3$ (input qubits at $-1, +1$ and ancillary qubits at $+i,-i$), can be used to find the period of an integer function with periodicity $r\leq 4$. In the special case where $r|4$ (i.e. $r\neq 3$), the test functions may be taken as $f_2(0)=f_2(2)=1, f_2(1)=f_2(3)=3$ and $f_4(0)=1, f_4(1) = 2, f_4(2) = 0, f_4(3) = 3$. For notational convenience $|\sigma_1\sigma_2\rangle$ is written $|2\sigma_1+\sigma_2\rangle$. The algorithm involves applications of the quantum Fourier transform which is typically entangling (Fig.~\ref{qft_fig}), producing flow fields with zeros that do not form equiangular halos (purple circles). The presence of equiangular halo zeros (blue and red circles) in certain states indicates separability or factorizability. In the final state (column 4), the input qubits are in a superposition of states $|k2^n/r\rangle$ for integer $k$. An additional measurement of the $n=2$ input qubits therefore gives the result $|k2^n/r\rangle$ for a specific value $k$, allowing $r$ to be deduced from multiple runs of the algorithm. Scale bar: length~1.} 
	\label{shor_fig}
\end{figure*}

This effect is also apparent in the quantum Fourier transform (Fig.~\ref{qft_fig}), an important operation in Shor's algorithm~\cite{shor1999polynomial} (Fig.~\ref{shor_fig}) amongst other applications~\cite{CleveR_Proceedings454_1998}. In a flow field representation with charge $d$, an individual qubit in a separable state is mapped to a factor of the overall complex function
\begin{align*}
    \alpha|0\rangle + \beta |1\rangle \mapsto \left(z-a_j\right)^{-d}\left(\alpha + \beta (z-a_j)^{2d} \right)
\end{align*}
where $a_j$ is the position of the $j$'th basis defect. The zeros of this function satisfy  $(z-a_j)^{2d} = -\alpha/\beta$, forming a regular $2d$-gon around $a_j$ (Fig.~\ref{qft_fig}a). Furthermore, as a separable state is continuously deformed into an entangled state, the deformation of the regular defect polygons (Fig.~\ref{qft_fig}b) demonstrates the emergence of entanglement.

\par
Defect dynamics in the flow field picture allow for the visualization of general quantum operations, including the period finding routine in Shor's algorithm~\cite{shor1999polynomial} (Fig.~\ref{shor_fig}). The defect halo criterion for separability makes the intermediate entangled states particularly apparent (Fig.~\ref{shor_fig}). The partial presence of defect halos (Fig.~\ref{shor_fig} second column, top row) additionally provides a necessary condition for the existence of a factorization of the underlying $n$-qubit state [SI]. The behavior of flow field defects thus captures the propagation of information in a variety of quantum processes, from qubit logic gates [SI] and the Deutsch-Josza algorithm~\cite{1997DeutschJozsa} [SI] to quantum Fourier transforms and Shor's algorithm (Fig.~\ref{shor_fig}).

\par

To complete the analogy with quantum processes, an inner product on the space of $n$-qubit vector fields, $\Phi^{(n)}\left(\mathcal{H}^{(n)}\right) \subset \mathcal {V}$, can be constructed by mapping the vector fields to $\mathbb{C}^{2^n}$. For some $\alpha \in \mathbb{C}$, define the map $\pi :\Phi^{(n)}\left(\mathcal{H}^{(n)}\right) \rightarrow \mathbb{C}^{2^n}$
\begin{align*}
f \mapsto \boldsymbol{v} = \left[ f(\alpha), Df(\alpha), D^2f(\alpha),..., D^{2^n-1}f(\alpha)  \right]
\end{align*}
where $f\in \Phi^{(n)}\left(\mathcal{H}^{(n)}\right)$ and $D$ denotes the complex derivative. An inner product on $\mathbb{C}^{2^n}$ gives rise to an inner product on $\mathcal{V}$, since $\pi$ is a linear map. We require the $n$-qubit basis states to be orthonormal under this inner product. Let $\boldsymbol{v}_{\boldsymbol{\sigma}}$ be the image of the vector field representing $|\boldsymbol{\sigma}\rangle$ under the map $\pi$, where $\boldsymbol{\sigma}$ is an $n$-bit binary string. We therefore need an inner product under which the vectors $\boldsymbol{v}_{\boldsymbol{\sigma}}$ are orthonormal. Let $B$ be the $2^n \times 2^n$ matrix whose columns are the vectors $\boldsymbol{v}_{\boldsymbol{\sigma}}$. Provided the $n$-qubit vector fields are linearly independent [SI], it is always possible to choose the evaluation point $\alpha \in \mathbb{C}$ so that $B$ is invertible~\cite{bocher1900theory}. Finally, $P = (B^{-1})^\dagger B^{-1}$, is the required inner product, satisfying $\langle \boldsymbol{v}_{\boldsymbol{\sigma}}, \boldsymbol{v}_{\boldsymbol{\sigma}'} \rangle_P = \boldsymbol{v}_{\boldsymbol{\sigma}}^\dagger P \boldsymbol{v}_{\boldsymbol{\sigma}'} = \delta_{\boldsymbol{\sigma}\boldsymbol{\sigma}'}$. For functions $f_1,f_2\in \Phi^{(n)}\left(\mathcal{H}^{(n)}\right)$, this becomes $\langle f_1, f_2 \rangle = \langle \pi(f_1), \pi(f_2) \rangle_P
$.

\par

To conclude, the mappings discussed here demonstrate the Hilbert space structure of 2D flow fields. By capturing both the linear structure and the tensor product structure, such flow fields provide possible classical  representations of entangled qubit states. The underlying construction is general:  different choices of the linear maps $\phi_j$ from a single qubit Hilbert space to 2D vector fields, will yield further realizations [SI]. Recent technological advances in the control of topological structures~\cite{MachonT_ProcRSocA472_2016} in fluids~\cite{ScheelerMW_Science357_2017} and other soft systems~\cite{TranL_ProcNatlAcadSci113_2016,ElgetiJ_SoftMatter7_2011,PengC_Science354_2016,DoostmohammadiA_NatCommun7_2016}, are promising candidates for experimentally realizing some of these vector field structures in two-dimensional geometries~\cite{TangX_SoftMatter13_2017, KeberFC_Science345_2014}. 

The observation that infinitely many alternative representations exist [SI] suggests an opportunity for tailoring representations towards physical robustness and optimal scalability. From a theoretical perspective, the qubit-flow analogy presents another avenue for studying topological defects. For example, the topology associated with qubit Hilbert spaces, such as the 2-qubit Hopf fibration~\cite{Mosseri_2001}, can be mapped to the flow field picture. Conversely,  analysis of the corresponding classical flow fields may provide insight into the topology of more complex  qubit systems. 
The qubit analogy also presents an opportunity to extend a previously noted correspondence between quantum theory and classical statistical mechanics~\cite{ChandlerD_JChemPhys74_1981}. In particular, the mappings presented here raise the interesting question of how defect entanglement and entanglement entropy~\cite{deffner2016foundations,sone2021quantum} could be generalized to flow and deformation fields in more complex fluids and continuum systems, including passive~\cite{machon2014knotted} and active~\cite{duclos2020topological} liquid crystals and  membranes~\cite{liu2021topological}. 

\par

We thank Bjorn Poonen for suggesting the linear independence argument [SI], and Aaron Levin for bringing to our attention Ref.~\cite{green1975some}. We are grateful to Martin Zwierlein and Vili Heinonen for many insightful comments.  This work was supported by a MathWorks fellowship (V.P.P.),  a Stanford Science fellowship (V.P.P.), by the Slovenian Research Agency (ARRS) under contracts P1-0099 and N1-0124 (\v{Z}.K.), and by the Robert E. Collins Distinguished Scholarship Fund (J.D.).


\bibliography{qbit_references}

\begin{thebibliography}{49}%
\makeatletter
\providecommand \@ifxundefined [1]{%
 \@ifx{#1\undefined}
}%
\providecommand \@ifnum [1]{%
 \ifnum #1\expandafter \@firstoftwo
 \else \expandafter \@secondoftwo
 \fi
}%
\providecommand \@ifx [1]{%
 \ifx #1\expandafter \@firstoftwo
 \else \expandafter \@secondoftwo
 \fi
}%
\providecommand \natexlab [1]{#1}%
\providecommand \enquote  [1]{``#1''}%
\providecommand \bibnamefont  [1]{#1}%
\providecommand \bibfnamefont [1]{#1}%
\providecommand \citenamefont [1]{#1}%
\providecommand \href@noop [0]{\@secondoftwo}%
\providecommand \href [0]{\begingroup \@sanitize@url \@href}%
\providecommand \@href[1]{\@@startlink{#1}\@@href}%
\providecommand \@@href[1]{\endgroup#1\@@endlink}%
\providecommand \@sanitize@url [0]{\catcode `\\12\catcode `\$12\catcode
  `\&12\catcode `\#12\catcode `\^12\catcode `\_12\catcode `\%12\relax}%
\providecommand \@@startlink[1]{}%
\providecommand \@@endlink[0]{}%
\providecommand \url  [0]{\begingroup\@sanitize@url \@url }%
\providecommand \@url [1]{\endgroup\@href {#1}{\urlprefix }}%
\providecommand \urlprefix  [0]{URL }%
\providecommand \Eprint [0]{\href }%
\providecommand \doibase [0]{https://doi.org/}%
\providecommand \selectlanguage [0]{\@gobble}%
\providecommand \bibinfo  [0]{\@secondoftwo}%
\providecommand \bibfield  [0]{\@secondoftwo}%
\providecommand \translation [1]{[#1]}%
\providecommand \BibitemOpen [0]{}%
\providecommand \bibitemStop [0]{}%
\providecommand \bibitemNoStop [0]{.\EOS\space}%
\providecommand \EOS [0]{\spacefactor3000\relax}%
\providecommand \BibitemShut  [1]{\csname bibitem#1\endcsname}%
\let\auto@bib@innerbib\@empty
\bibitem [{\citenamefont {Prakash}\ and\ \citenamefont
  {Gershenfeld}(2007)}]{prakash2007microfluidic}%
  \BibitemOpen
  \bibfield  {author} {\bibinfo {author} {\bibfnamefont {M.}~\bibnamefont
  {Prakash}}\ and\ \bibinfo {author} {\bibfnamefont {N.}~\bibnamefont
  {Gershenfeld}},\ }\bibfield  {title} {\bibinfo {title} {Microfluidic bubble
  logic},\ }\href@noop {} {\bibfield  {journal} {\bibinfo  {journal} {Science}\
  }\textbf {\bibinfo {volume} {315}},\ \bibinfo {pages} {832} (\bibinfo {year}
  {2007})}\BibitemShut {NoStop}%
\bibitem [{\citenamefont {Gilpin}(2018)}]{gilpin2018cryptographic}%
  \BibitemOpen
  \bibfield  {author} {\bibinfo {author} {\bibfnamefont {W.}~\bibnamefont
  {Gilpin}},\ }\bibfield  {title} {\bibinfo {title} {Cryptographic hashing
  using chaotic hydrodynamics},\ }\href@noop {} {\bibfield  {journal} {\bibinfo
   {journal} {Proc. Natl. Acad. Sci. U.S.A.}\ }\textbf {\bibinfo {volume}
  {115}},\ \bibinfo {pages} {4869} (\bibinfo {year} {2018})}\BibitemShut
  {NoStop}%
\bibitem [{\citenamefont {Chiu}\ \emph {et~al.}(2001)\citenamefont {Chiu},
  \citenamefont {Pezzoli}, \citenamefont {Wu}, \citenamefont {Stroock},\ and\
  \citenamefont {Whitesides}}]{ChiuDT_ProcNatlAcadSci98_2001}%
  \BibitemOpen
  \bibfield  {author} {\bibinfo {author} {\bibfnamefont {D.~T.}\ \bibnamefont
  {Chiu}}, \bibinfo {author} {\bibfnamefont {E.}~\bibnamefont {Pezzoli}},
  \bibinfo {author} {\bibfnamefont {H.}~\bibnamefont {Wu}}, \bibinfo {author}
  {\bibfnamefont {A.~D.}\ \bibnamefont {Stroock}},\ and\ \bibinfo {author}
  {\bibfnamefont {G.~M.}\ \bibnamefont {Whitesides}},\ }\bibfield  {title}
  {\bibinfo {title} {Using three-dimensional microfluidic networks for solving
  computationally hard problems},\ }\href
  {http://dx.doi.org/10.1073/pnas.061014198} {\bibfield  {journal} {\bibinfo
  {journal} {Proc. Natl. Acad. Sci. U.S.A.}\ }\textbf {\bibinfo {volume}
  {98}},\ \bibinfo {pages} {2961} (\bibinfo {year} {2001})}\BibitemShut
  {NoStop}%
\bibitem [{\citenamefont {Gunji}\ \emph {et~al.}(2011)\citenamefont {Gunji},
  \citenamefont {Nishiyama},\ and\ \citenamefont
  {Adamatzky}}]{gunji2011robust}%
  \BibitemOpen
  \bibfield  {author} {\bibinfo {author} {\bibfnamefont {Y.-P.}\ \bibnamefont
  {Gunji}}, \bibinfo {author} {\bibfnamefont {Y.}~\bibnamefont {Nishiyama}},\
  and\ \bibinfo {author} {\bibfnamefont {A.}~\bibnamefont {Adamatzky}},\
  }\bibfield  {title} {\bibinfo {title} {Robust soldier crab ball gate},\
  }\href@noop {} {\bibfield  {journal} {\bibinfo  {journal} {Complex Syst.}\
  }\textbf {\bibinfo {volume} {20}},\ \bibinfo {pages} {93} (\bibinfo {year}
  {2011})}\BibitemShut {NoStop}%
\bibitem [{\citenamefont {Woodhouse}\ and\ \citenamefont
  {Dunkel}(2017)}]{woodhouse2017active}%
  \BibitemOpen
  \bibfield  {author} {\bibinfo {author} {\bibfnamefont {F.~G.}\ \bibnamefont
  {Woodhouse}}\ and\ \bibinfo {author} {\bibfnamefont {J.}~\bibnamefont
  {Dunkel}},\ }\bibfield  {title} {\bibinfo {title} {Active matter logic for
  autonomous microfluidics},\ }\href@noop {} {\bibfield  {journal} {\bibinfo
  {journal} {Nat. Commun.}\ }\textbf {\bibinfo {volume} {8}},\ \bibinfo {pages}
  {1} (\bibinfo {year} {2017})}\BibitemShut {NoStop}%
\bibitem [{\citenamefont {Nicolau}\ \emph {et~al.}(2016)\citenamefont
  {Nicolau}, \citenamefont {Lard}, \citenamefont {Korten}, \citenamefont {van
  Delft}, \citenamefont {Persson}, \citenamefont {Bengtsson}, \citenamefont
  {Månsson}, \citenamefont {Diez}, \citenamefont {Linke},\ and\ \citenamefont
  {Nicolau}}]{NicolauDV_ProcNatlAcadSci113_2016}%
  \BibitemOpen
  \bibfield  {author} {\bibinfo {author} {\bibfnamefont {D.~V.}\ \bibnamefont
  {Nicolau}}, \bibinfo {author} {\bibfnamefont {M.}~\bibnamefont {Lard}},
  \bibinfo {author} {\bibfnamefont {T.}~\bibnamefont {Korten}}, \bibinfo
  {author} {\bibfnamefont {F.~C. M. J.~M.}\ \bibnamefont {van Delft}}, \bibinfo
  {author} {\bibfnamefont {M.}~\bibnamefont {Persson}}, \bibinfo {author}
  {\bibfnamefont {E.}~\bibnamefont {Bengtsson}}, \bibinfo {author}
  {\bibfnamefont {A.}~\bibnamefont {Månsson}}, \bibinfo {author}
  {\bibfnamefont {S.}~\bibnamefont {Diez}}, \bibinfo {author} {\bibfnamefont
  {H.}~\bibnamefont {Linke}},\ and\ \bibinfo {author} {\bibfnamefont {D.~V.}\
  \bibnamefont {Nicolau}},\ }\bibfield  {title} {\bibinfo {title} {Parallel
  computation with molecular-motor-propelled agents in nanofabricated
  networks},\ }\href {http://dx.doi.org/10.1073/pnas.1510825113} {\bibfield
  {journal} {\bibinfo  {journal} {Proc. Natl. Acad. Sci. U.S.A.}\ }\textbf
  {\bibinfo {volume} {113}},\ \bibinfo {pages} {2591} (\bibinfo {year}
  {2016})}\BibitemShut {NoStop}%
\bibitem [{\citenamefont {Qian}\ and\ \citenamefont
  {Winfree}(2011)}]{qian2011scaling}%
  \BibitemOpen
  \bibfield  {author} {\bibinfo {author} {\bibfnamefont {L.}~\bibnamefont
  {Qian}}\ and\ \bibinfo {author} {\bibfnamefont {E.}~\bibnamefont {Winfree}},\
  }\bibfield  {title} {\bibinfo {title} {Scaling up digital circuit computation
  with dna strand displacement cascades},\ }\href@noop {} {\bibfield  {journal}
  {\bibinfo  {journal} {Science}\ }\textbf {\bibinfo {volume} {332}},\ \bibinfo
  {pages} {1196} (\bibinfo {year} {2011})}\BibitemShut {NoStop}%
\bibitem [{\citenamefont {Fruchart}\ \emph {et~al.}(2020)\citenamefont
  {Fruchart}, \citenamefont {Zhou},\ and\ \citenamefont
  {Vitelli}}]{fruchart2020dualities}%
  \BibitemOpen
  \bibfield  {author} {\bibinfo {author} {\bibfnamefont {M.}~\bibnamefont
  {Fruchart}}, \bibinfo {author} {\bibfnamefont {Y.}~\bibnamefont {Zhou}},\
  and\ \bibinfo {author} {\bibfnamefont {V.}~\bibnamefont {Vitelli}},\
  }\bibfield  {title} {\bibinfo {title} {Dualities and non-abelian mechanics},\
  }\href@noop {} {\bibfield  {journal} {\bibinfo  {journal} {Nature}\ }\textbf
  {\bibinfo {volume} {577}},\ \bibinfo {pages} {636} (\bibinfo {year}
  {2020})}\BibitemShut {NoStop}%
\bibitem [{\citenamefont {Ladd}\ \emph {et~al.}(2010)\citenamefont {Ladd},
  \citenamefont {Jelezko}, \citenamefont {Laflamme}, \citenamefont {Nakamura},
  \citenamefont {Monroe},\ and\ \citenamefont {O'Brien}}]{Ladd:2010aa}%
  \BibitemOpen
  \bibfield  {author} {\bibinfo {author} {\bibfnamefont {T.~D.}\ \bibnamefont
  {Ladd}}, \bibinfo {author} {\bibfnamefont {F.}~\bibnamefont {Jelezko}},
  \bibinfo {author} {\bibfnamefont {R.}~\bibnamefont {Laflamme}}, \bibinfo
  {author} {\bibfnamefont {Y.}~\bibnamefont {Nakamura}}, \bibinfo {author}
  {\bibfnamefont {C.}~\bibnamefont {Monroe}},\ and\ \bibinfo {author}
  {\bibfnamefont {J.~L.}\ \bibnamefont {O'Brien}},\ }\bibfield  {title}
  {\bibinfo {title} {Quantum computers},\ }\href
  {https://doi.org/10.1038/nature08812} {\bibfield  {journal} {\bibinfo
  {journal} {Nature}\ }\textbf {\bibinfo {volume} {464}},\ \bibinfo {pages}
  {45} (\bibinfo {year} {2010})}\BibitemShut {NoStop}%
\bibitem [{\citenamefont {\v{Z}. Kos}\ and\ \citenamefont
  {Dunkel}(2020)}]{nematic_bits}%
  \BibitemOpen
  \bibfield  {author} {\bibinfo {author} {\bibnamefont {\v{Z}. Kos}}\ and\
  \bibinfo {author} {\bibfnamefont {J.}~\bibnamefont {Dunkel}},\ }\href@noop {}
  {\bibinfo {title} {Nematic bits and logic gates}} (\bibinfo {year} {2020}),\
  \Eprint {https://arxiv.org/abs/2008.13094} {arXiv:2008.13094 [cond-mat.soft]}
  \BibitemShut {NoStop}%
\bibitem [{\citenamefont {Karimi}\ and\ \citenamefont
  {Boyd}(2015)}]{karimi2015classical}%
  \BibitemOpen
  \bibfield  {author} {\bibinfo {author} {\bibfnamefont {E.}~\bibnamefont
  {Karimi}}\ and\ \bibinfo {author} {\bibfnamefont {R.~W.}\ \bibnamefont
  {Boyd}},\ }\bibfield  {title} {\bibinfo {title} {Classical entanglement?},\
  }\href@noop {} {\bibfield  {journal} {\bibinfo  {journal} {Science}\ }\textbf
  {\bibinfo {volume} {350}},\ \bibinfo {pages} {1172} (\bibinfo {year}
  {2015})}\BibitemShut {NoStop}%
\bibitem [{\citenamefont {Spreeuw}(1998)}]{spreeuw1998classical}%
  \BibitemOpen
  \bibfield  {author} {\bibinfo {author} {\bibfnamefont {R.~J.}\ \bibnamefont
  {Spreeuw}},\ }\bibfield  {title} {\bibinfo {title} {A classical analogy of
  entanglement},\ }\href@noop {} {\bibfield  {journal} {\bibinfo  {journal}
  {Found. Phys.}\ }\textbf {\bibinfo {volume} {28}},\ \bibinfo {pages} {361}
  (\bibinfo {year} {1998})}\BibitemShut {NoStop}%
\bibitem [{\citenamefont {Goldin}\ \emph {et~al.}(2010)\citenamefont {Goldin},
  \citenamefont {Francisco},\ and\ \citenamefont
  {Ledesma}}]{goldin2010simulating}%
  \BibitemOpen
  \bibfield  {author} {\bibinfo {author} {\bibfnamefont {M.~A.}\ \bibnamefont
  {Goldin}}, \bibinfo {author} {\bibfnamefont {D.}~\bibnamefont {Francisco}},\
  and\ \bibinfo {author} {\bibfnamefont {S.}~\bibnamefont {Ledesma}},\
  }\bibfield  {title} {\bibinfo {title} {Simulating bell inequality violations
  with classical optics encoded qubits},\ }\href@noop {} {\bibfield  {journal}
  {\bibinfo  {journal} {J. Opt. Soc. Am.}\ }\textbf {\bibinfo {volume} {27}},\
  \bibinfo {pages} {779} (\bibinfo {year} {2010})}\BibitemShut {NoStop}%
\bibitem [{\citenamefont {Song}\ \emph {et~al.}(2015)\citenamefont {Song},
  \citenamefont {Sun}, \citenamefont {Li}, \citenamefont {Qin},\ and\
  \citenamefont {Zhang}}]{song2015bell}%
  \BibitemOpen
  \bibfield  {author} {\bibinfo {author} {\bibfnamefont {X.}~\bibnamefont
  {Song}}, \bibinfo {author} {\bibfnamefont {Y.}~\bibnamefont {Sun}}, \bibinfo
  {author} {\bibfnamefont {P.}~\bibnamefont {Li}}, \bibinfo {author}
  {\bibfnamefont {H.}~\bibnamefont {Qin}},\ and\ \bibinfo {author}
  {\bibfnamefont {X.}~\bibnamefont {Zhang}},\ }\bibfield  {title} {\bibinfo
  {title} {Bell’s measure and implementing quantum fourier transform with
  orbital angular momentum of classical light},\ }\href@noop {} {\bibfield
  {journal} {\bibinfo  {journal} {Sci. Rep.}\ }\textbf {\bibinfo {volume}
  {5}},\ \bibinfo {pages} {1} (\bibinfo {year} {2015})}\BibitemShut {NoStop}%
\bibitem [{\citenamefont {Kedia}\ \emph {et~al.}(2013)\citenamefont {Kedia},
  \citenamefont {Bialynicki-Birula}, \citenamefont {Peralta-Salas},\ and\
  \citenamefont {Irvine}}]{kedia2013tying}%
  \BibitemOpen
  \bibfield  {author} {\bibinfo {author} {\bibfnamefont {H.}~\bibnamefont
  {Kedia}}, \bibinfo {author} {\bibfnamefont {I.}~\bibnamefont
  {Bialynicki-Birula}}, \bibinfo {author} {\bibfnamefont {D.}~\bibnamefont
  {Peralta-Salas}},\ and\ \bibinfo {author} {\bibfnamefont {W.~T.}\
  \bibnamefont {Irvine}},\ }\bibfield  {title} {\bibinfo {title} {Tying knots
  in light fields},\ }\href@noop {} {\bibfield  {journal} {\bibinfo  {journal}
  {Phys. Rev. Lett.}\ }\textbf {\bibinfo {volume} {111}},\ \bibinfo {pages}
  {150404} (\bibinfo {year} {2013})}\BibitemShut {NoStop}%
\bibitem [{\citenamefont {Brown}(2020)}]{BrownAR_Quantum4_2020}%
  \BibitemOpen
  \bibfield  {author} {\bibinfo {author} {\bibfnamefont {A.~R.}\ \bibnamefont
  {Brown}},\ }\bibfield  {title} {\bibinfo {title} {Playing {Pool} with
  {$|\psi\rangle$} : from {Bouncing} {Billiards} to {Quantum} {Search}},\
  }\href {http://dx.doi.org/10.22331/q-2020-11-02-357} {\bibfield  {journal}
  {\bibinfo  {journal} {Quantum}\ }\textbf {\bibinfo {volume} {4}},\ \bibinfo
  {pages} {357} (\bibinfo {year} {2020})}\BibitemShut {NoStop}%
\bibitem [{\citenamefont {Bush}\ and\ \citenamefont
  {Oza}(2020)}]{BushJWM_RepProgPhys84_2020}%
  \BibitemOpen
  \bibfield  {author} {\bibinfo {author} {\bibfnamefont {J.~W.~M.}\
  \bibnamefont {Bush}}\ and\ \bibinfo {author} {\bibfnamefont {A.~U.}\
  \bibnamefont {Oza}},\ }\bibfield  {title} {\bibinfo {title} {Hydrodynamic
  quantum analogs},\ }\href {http://dx.doi.org/10.1088/1361-6633/abc22c}
  {\bibfield  {journal} {\bibinfo  {journal} {Rep. Prog. Phys.}\ }\textbf
  {\bibinfo {volume} {84}},\ \bibinfo {pages} {017001} (\bibinfo {year}
  {2020})}\BibitemShut {NoStop}%
\bibitem [{\citenamefont {Bush}\ \emph {et~al.}(2018)\citenamefont {Bush},
  \citenamefont {Couder}, \citenamefont {Gilet}, \citenamefont {Milewski},\
  and\ \citenamefont {Nachbin}}]{BushJWM_Chaos28_2018}%
  \BibitemOpen
  \bibfield  {author} {\bibinfo {author} {\bibfnamefont {J.~W.~M.}\
  \bibnamefont {Bush}}, \bibinfo {author} {\bibfnamefont {Y.}~\bibnamefont
  {Couder}}, \bibinfo {author} {\bibfnamefont {T.}~\bibnamefont {Gilet}},
  \bibinfo {author} {\bibfnamefont {P.~A.}\ \bibnamefont {Milewski}},\ and\
  \bibinfo {author} {\bibfnamefont {A.}~\bibnamefont {Nachbin}},\ }\bibfield
  {title} {\bibinfo {title} {Introduction to focus issue on hydrodynamic
  quantum analogs},\ }\href {http://dx.doi.org/10.1063/1.5055383} {\bibfield
  {journal} {\bibinfo  {journal} {Chaos}\ }\textbf {\bibinfo {volume} {28}},\
  \bibinfo {pages} {096001} (\bibinfo {year} {2018})}\BibitemShut {NoStop}%
\bibitem [{\citenamefont {M{\"a}kel{\"a}}\ and\ \citenamefont
  {Messina}(2010{\natexlab{a}})}]{makela2010n}%
  \BibitemOpen
  \bibfield  {author} {\bibinfo {author} {\bibfnamefont {H.}~\bibnamefont
  {M{\"a}kel{\"a}}}\ and\ \bibinfo {author} {\bibfnamefont {A.}~\bibnamefont
  {Messina}},\ }\bibfield  {title} {\bibinfo {title} {N-qubit states as points
  on the bloch sphere},\ }\href@noop {} {\bibfield  {journal} {\bibinfo
  {journal} {Phys. Scr.}\ }\textbf {\bibinfo {volume} {2010}},\ \bibinfo
  {pages} {014054} (\bibinfo {year} {2010}{\natexlab{a}})}\BibitemShut
  {NoStop}%
\bibitem [{\citenamefont {M{\"a}kel{\"a}}\ and\ \citenamefont
  {Messina}(2010{\natexlab{b}})}]{makela2010polynomial}%
  \BibitemOpen
  \bibfield  {author} {\bibinfo {author} {\bibfnamefont {H.}~\bibnamefont
  {M{\"a}kel{\"a}}}\ and\ \bibinfo {author} {\bibfnamefont {A.}~\bibnamefont
  {Messina}},\ }\bibfield  {title} {\bibinfo {title} {Polynomial method to
  study the entanglement of pure n-qubit states},\ }\href@noop {} {\bibfield
  {journal} {\bibinfo  {journal} {Phys. Rev. A}\ }\textbf {\bibinfo {volume}
  {81}},\ \bibinfo {pages} {012326} (\bibinfo {year}
  {2010}{\natexlab{b}})}\BibitemShut {NoStop}%
\bibitem [{\citenamefont {Bruno}(2012)}]{bruno2012quantum}%
  \BibitemOpen
  \bibfield  {author} {\bibinfo {author} {\bibfnamefont {P.}~\bibnamefont
  {Bruno}},\ }\bibfield  {title} {\bibinfo {title} {Quantum geometric phase in
  majorana’s stellar representation: mapping onto a many-body aharonov-bohm
  phase},\ }\href@noop {} {\bibfield  {journal} {\bibinfo  {journal} {Phys.
  Rev. Lett.}\ }\textbf {\bibinfo {volume} {108}},\ \bibinfo {pages} {240402}
  (\bibinfo {year} {2012})}\BibitemShut {NoStop}%
\bibitem [{\citenamefont {Kam}\ and\ \citenamefont {Liu}(2020)}]{kam2020three}%
  \BibitemOpen
  \bibfield  {author} {\bibinfo {author} {\bibfnamefont {C.-F.}\ \bibnamefont
  {Kam}}\ and\ \bibinfo {author} {\bibfnamefont {R.-B.}\ \bibnamefont {Liu}},\
  }\bibfield  {title} {\bibinfo {title} {Three-tangle of a general three-qubit
  state in the representation of majorana stars},\ }\href@noop {} {\bibfield
  {journal} {\bibinfo  {journal} {Phys. Rev. A}\ }\textbf {\bibinfo {volume}
  {101}},\ \bibinfo {pages} {032318} (\bibinfo {year} {2020})}\BibitemShut
  {NoStop}%
\bibitem [{\citenamefont {Devi}\ and\ \citenamefont
  {Rajagopal}(2012)}]{devi2012majorana}%
  \BibitemOpen
  \bibfield  {author} {\bibinfo {author} {\bibfnamefont {A.~U.}\ \bibnamefont
  {Devi}}\ and\ \bibinfo {author} {\bibfnamefont {A.}~\bibnamefont
  {Rajagopal}},\ }\bibfield  {title} {\bibinfo {title} {Majorana representation
  of symmetric multiqubit states},\ }\href@noop {} {\bibfield  {journal}
  {\bibinfo  {journal} {Quantum Inf. Process.}\ }\textbf {\bibinfo {volume}
  {11}},\ \bibinfo {pages} {685} (\bibinfo {year} {2012})}\BibitemShut
  {NoStop}%
\bibitem [{\citenamefont {Serrano-Ens{\'a}stiga}\ and\ \citenamefont
  {Braun}(2020)}]{serrano2020majorana}%
  \BibitemOpen
  \bibfield  {author} {\bibinfo {author} {\bibfnamefont {E.}~\bibnamefont
  {Serrano-Ens{\'a}stiga}}\ and\ \bibinfo {author} {\bibfnamefont
  {D.}~\bibnamefont {Braun}},\ }\bibfield  {title} {\bibinfo {title} {Majorana
  representation for mixed states},\ }\href@noop {} {\bibfield  {journal}
  {\bibinfo  {journal} {Phys. Rev. A}\ }\textbf {\bibinfo {volume} {101}},\
  \bibinfo {pages} {022332} (\bibinfo {year} {2020})}\BibitemShut {NoStop}%
\bibitem [{\citenamefont {Crowdy}(2010)}]{crowdy2010new}%
  \BibitemOpen
  \bibfield  {author} {\bibinfo {author} {\bibfnamefont {D.}~\bibnamefont
  {Crowdy}},\ }\bibfield  {title} {\bibinfo {title} {A new calculus for
  two-dimensional vortex dynamics},\ }\href@noop {} {\bibfield  {journal}
  {\bibinfo  {journal} {Theor. Comput. Fluid Dyn.}\ }\textbf {\bibinfo {volume}
  {24}},\ \bibinfo {pages} {9} (\bibinfo {year} {2010})}\BibitemShut {NoStop}%
\bibitem [{\citenamefont {Cleve}\ \emph {et~al.}(1998)\citenamefont {Cleve},
  \citenamefont {Ekert}, \citenamefont {Macchiavello},\ and\ \citenamefont
  {Mosca}}]{CleveR_Proceedings454_1998}%
  \BibitemOpen
  \bibfield  {author} {\bibinfo {author} {\bibfnamefont {R.}~\bibnamefont
  {Cleve}}, \bibinfo {author} {\bibfnamefont {A.}~\bibnamefont {Ekert}},
  \bibinfo {author} {\bibfnamefont {C.}~\bibnamefont {Macchiavello}},\ and\
  \bibinfo {author} {\bibfnamefont {M.}~\bibnamefont {Mosca}},\ }\bibfield
  {title} {\bibinfo {title} {Quantum algorithms revisited},\ }\href
  {http://dx.doi.org/10.1098/rspa.1998.0164} {\bibfield  {journal} {\bibinfo
  {journal} {Proc. R. Soc. Lond. A}\ }\textbf {\bibinfo {volume} {454}},\
  \bibinfo {pages} {339} (\bibinfo {year} {1998})}\BibitemShut {NoStop}%
\bibitem [{\citenamefont {Shor}(1997)}]{shor1999polynomial}%
  \BibitemOpen
  \bibfield  {author} {\bibinfo {author} {\bibfnamefont {P.~W.}\ \bibnamefont
  {Shor}},\ }\bibfield  {title} {\bibinfo {title} {Polynomial-time algorithms
  for prime factorization and discrete logarithms on a quantum computer},\
  }\href@noop {} {\bibfield  {journal} {\bibinfo  {journal} {SIAM J. Comput.}\
  }\textbf {\bibinfo {volume} {26}},\ \bibinfo {pages} {1484} (\bibinfo {year}
  {1997})}\BibitemShut {NoStop}%
\bibitem [{\citenamefont {Furman}(1994)}]{furman1994compact}%
  \BibitemOpen
  \bibfield  {author} {\bibinfo {author} {\bibfnamefont {M.~A.}\ \bibnamefont
  {Furman}},\ }\bibfield  {title} {\bibinfo {title} {Compact complex
  expressions for the electric field of two-dimensional elliptical charge
  distributions},\ }\href@noop {} {\bibfield  {journal} {\bibinfo  {journal}
  {Am. J. Phys.}\ }\textbf {\bibinfo {volume} {62}},\ \bibinfo {pages} {1134}
  (\bibinfo {year} {1994})}\BibitemShut {NoStop}%
\bibitem [{\citenamefont {Khoromskaia}\ and\ \citenamefont
  {Alexander}(2017)}]{khoromskaia2017vortex}%
  \BibitemOpen
  \bibfield  {author} {\bibinfo {author} {\bibfnamefont {D.}~\bibnamefont
  {Khoromskaia}}\ and\ \bibinfo {author} {\bibfnamefont {G.~P.}\ \bibnamefont
  {Alexander}},\ }\bibfield  {title} {\bibinfo {title} {Vortex formation and
  dynamics of defects in active nematic shells},\ }\href@noop {} {\bibfield
  {journal} {\bibinfo  {journal} {New J. Phys.}\ }\textbf {\bibinfo {volume}
  {19}},\ \bibinfo {pages} {103043} (\bibinfo {year} {2017})}\BibitemShut
  {NoStop}%
\bibitem [{\citenamefont {Tang}\ and\ \citenamefont
  {Selinger}(2017)}]{TangX_SoftMatter13_2017}%
  \BibitemOpen
  \bibfield  {author} {\bibinfo {author} {\bibfnamefont {X.}~\bibnamefont
  {Tang}}\ and\ \bibinfo {author} {\bibfnamefont {J.~V.}\ \bibnamefont
  {Selinger}},\ }\bibfield  {title} {\bibinfo {title} {Orientation of
  topological defects in 2d nematic liquid crystals},\ }\href
  {http://dx.doi.org/10.1039/c7sm01195d} {\bibfield  {journal} {\bibinfo
  {journal} {Soft Matter}\ }\textbf {\bibinfo {volume} {13}},\ \bibinfo {pages}
  {5481} (\bibinfo {year} {2017})}\BibitemShut {NoStop}%
\bibitem [{\citenamefont {Sarkar}\ \emph {et~al.}(2021)\citenamefont {Sarkar},
  \citenamefont {{\v{C}}ebron}, \citenamefont {Brojan},\ and\ \citenamefont
  {Ko{\v{s}}mrlj}}]{sarkar2021elastic}%
  \BibitemOpen
  \bibfield  {author} {\bibinfo {author} {\bibfnamefont {S.}~\bibnamefont
  {Sarkar}}, \bibinfo {author} {\bibfnamefont {M.}~\bibnamefont
  {{\v{C}}ebron}}, \bibinfo {author} {\bibfnamefont {M.}~\bibnamefont
  {Brojan}},\ and\ \bibinfo {author} {\bibfnamefont {A.}~\bibnamefont
  {Ko{\v{s}}mrlj}},\ }\bibfield  {title} {\bibinfo {title} {Elastic multipole
  method for describing deformation of infinite two-dimensional solids with
  circular inclusions},\ }\href@noop {} {\bibfield  {journal} {\bibinfo
  {journal} {Phys. Rev. E}\ }\textbf {\bibinfo {volume} {103}},\ \bibinfo
  {pages} {053003} (\bibinfo {year} {2021})}\BibitemShut {NoStop}%
\bibitem [{\citenamefont {Muskhelishvili}(1963)}]{muskhelishvili1963some}%
  \BibitemOpen
  \bibfield  {author} {\bibinfo {author} {\bibfnamefont {N.}~\bibnamefont
  {Muskhelishvili}},\ }\href@noop {} {\emph {\bibinfo {title} {Some basic
  problems of the mathematical theory of elasticity}}}\ (\bibinfo  {publisher}
  {Noordhoff},\ \bibinfo {address} {Groningen},\ \bibinfo {year}
  {1963})\BibitemShut {NoStop}%
\bibitem [{\citenamefont {Deutsch}\ and\ \citenamefont
  {Jozsa}(1997)}]{1997DeutschJozsa}%
  \BibitemOpen
  \bibfield  {author} {\bibinfo {author} {\bibfnamefont {D.}~\bibnamefont
  {Deutsch}}\ and\ \bibinfo {author} {\bibfnamefont {R.}~\bibnamefont
  {Jozsa}},\ }\bibfield  {title} {\bibinfo {title} {Rapid solution of problems
  by quantum computation},\ }\href@noop {} {\bibfield  {journal} {\bibinfo
  {journal} {Proc. R. Soc. Lond. A}\ }\textbf {\bibinfo {volume} {439}},\
  \bibinfo {pages} {553} (\bibinfo {year} {1997})}\BibitemShut {NoStop}%
\bibitem [{\citenamefont {B{\^o}cher}(1900)}]{bocher1900theory}%
  \BibitemOpen
  \bibfield  {author} {\bibinfo {author} {\bibfnamefont {M.}~\bibnamefont
  {B{\^o}cher}},\ }\bibfield  {title} {\bibinfo {title} {The theory of linear
  dependence},\ }\href@noop {} {\bibfield  {journal} {\bibinfo  {journal} {Ann.
  Math.}\ }\textbf {\bibinfo {volume} {2}},\ \bibinfo {pages} {81} (\bibinfo
  {year} {1900})}\BibitemShut {NoStop}%
\bibitem [{\citenamefont {Machon}\ and\ \citenamefont
  {Alexander}(2016)}]{MachonT_ProcRSocA472_2016}%
  \BibitemOpen
  \bibfield  {author} {\bibinfo {author} {\bibfnamefont {T.}~\bibnamefont
  {Machon}}\ and\ \bibinfo {author} {\bibfnamefont {G.~P.}\ \bibnamefont
  {Alexander}},\ }\bibfield  {title} {\bibinfo {title} {Global defect topology
  in nematic liquid crystals},\ }\href
  {http://dx.doi.org/10.1098/rspa.2016.0265} {\bibfield  {journal} {\bibinfo
  {journal} {Proc. R. Soc. Lond. A}\ }\textbf {\bibinfo {volume} {472}},\
  \bibinfo {pages} {20160265} (\bibinfo {year} {2016})}\BibitemShut {NoStop}%
\bibitem [{\citenamefont {Scheeler}\ \emph {et~al.}(2017)\citenamefont
  {Scheeler}, \citenamefont {van Rees}, \citenamefont {Kedia}, \citenamefont
  {Kleckner},\ and\ \citenamefont {Irvine}}]{ScheelerMW_Science357_2017}%
  \BibitemOpen
  \bibfield  {author} {\bibinfo {author} {\bibfnamefont {M.~W.}\ \bibnamefont
  {Scheeler}}, \bibinfo {author} {\bibfnamefont {W.~M.}\ \bibnamefont {van
  Rees}}, \bibinfo {author} {\bibfnamefont {H.}~\bibnamefont {Kedia}}, \bibinfo
  {author} {\bibfnamefont {D.}~\bibnamefont {Kleckner}},\ and\ \bibinfo
  {author} {\bibfnamefont {W.~T.~M.}\ \bibnamefont {Irvine}},\ }\bibfield
  {title} {\bibinfo {title} {Complete measurement of helicity and its dynamics
  in vortex tubes},\ }\href {http://dx.doi.org/10.1126/science.aam6897}
  {\bibfield  {journal} {\bibinfo  {journal} {Science}\ }\textbf {\bibinfo
  {volume} {357}},\ \bibinfo {pages} {487} (\bibinfo {year}
  {2017})}\BibitemShut {NoStop}%
\bibitem [{\citenamefont {Tran}\ \emph {et~al.}(2016)\citenamefont {Tran},
  \citenamefont {Lavrentovich}, \citenamefont {Beller}, \citenamefont {Li},
  \citenamefont {Stebe},\ and\ \citenamefont
  {Kamien}}]{TranL_ProcNatlAcadSci113_2016}%
  \BibitemOpen
  \bibfield  {author} {\bibinfo {author} {\bibfnamefont {L.}~\bibnamefont
  {Tran}}, \bibinfo {author} {\bibfnamefont {M.~O.}\ \bibnamefont
  {Lavrentovich}}, \bibinfo {author} {\bibfnamefont {D.~A.}\ \bibnamefont
  {Beller}}, \bibinfo {author} {\bibfnamefont {N.}~\bibnamefont {Li}}, \bibinfo
  {author} {\bibfnamefont {K.~J.}\ \bibnamefont {Stebe}},\ and\ \bibinfo
  {author} {\bibfnamefont {R.~D.}\ \bibnamefont {Kamien}},\ }\bibfield  {title}
  {\bibinfo {title} {Lassoing saddle splay and the geometrical control of
  topological defects},\ }\href {http://dx.doi.org/10.1073/pnas.1602703113}
  {\bibfield  {journal} {\bibinfo  {journal} {Proc. Natl. Acad. Sci. U.S.A.}\
  }\textbf {\bibinfo {volume} {113}},\ \bibinfo {pages} {7106} (\bibinfo {year}
  {2016})}\BibitemShut {NoStop}%
\bibitem [{\citenamefont {Elgeti}\ \emph {et~al.}(2011)\citenamefont {Elgeti},
  \citenamefont {Cates},\ and\ \citenamefont
  {Marenduzzo}}]{ElgetiJ_SoftMatter7_2011}%
  \BibitemOpen
  \bibfield  {author} {\bibinfo {author} {\bibfnamefont {J.}~\bibnamefont
  {Elgeti}}, \bibinfo {author} {\bibfnamefont {M.~E.}\ \bibnamefont {Cates}},\
  and\ \bibinfo {author} {\bibfnamefont {D.}~\bibnamefont {Marenduzzo}},\
  }\bibfield  {title} {\bibinfo {title} {Defect hydrodynamics in 2d polar
  active fluids},\ }\href {http://dx.doi.org/10.1039/c0sm01097a} {\bibfield
  {journal} {\bibinfo  {journal} {Soft Matter}\ }\textbf {\bibinfo {volume}
  {7}},\ \bibinfo {pages} {3177} (\bibinfo {year} {2011})}\BibitemShut
  {NoStop}%
\bibitem [{\citenamefont {Peng}\ \emph {et~al.}(2016)\citenamefont {Peng},
  \citenamefont {Turiv}, \citenamefont {Guo}, \citenamefont {Wei},\ and\
  \citenamefont {Lavrentovich}}]{PengC_Science354_2016}%
  \BibitemOpen
  \bibfield  {author} {\bibinfo {author} {\bibfnamefont {C.}~\bibnamefont
  {Peng}}, \bibinfo {author} {\bibfnamefont {T.}~\bibnamefont {Turiv}},
  \bibinfo {author} {\bibfnamefont {Y.}~\bibnamefont {Guo}}, \bibinfo {author}
  {\bibfnamefont {Q.-H.}\ \bibnamefont {Wei}},\ and\ \bibinfo {author}
  {\bibfnamefont {O.~D.}\ \bibnamefont {Lavrentovich}},\ }\bibfield  {title}
  {\bibinfo {title} {Command of active matter by topological defects and
  patterns},\ }\href {http://dx.doi.org/10.1126/science.aah6936} {\bibfield
  {journal} {\bibinfo  {journal} {Science}\ }\textbf {\bibinfo {volume}
  {354}},\ \bibinfo {pages} {882} (\bibinfo {year} {2016})}\BibitemShut
  {NoStop}%
\bibitem [{\citenamefont {Doostmohammadi}\ \emph {et~al.}(2016)\citenamefont
  {Doostmohammadi}, \citenamefont {Adamer}, \citenamefont {Thampi},\ and\
  \citenamefont {Yeomans}}]{DoostmohammadiA_NatCommun7_2016}%
  \BibitemOpen
  \bibfield  {author} {\bibinfo {author} {\bibfnamefont {A.}~\bibnamefont
  {Doostmohammadi}}, \bibinfo {author} {\bibfnamefont {M.~F.}\ \bibnamefont
  {Adamer}}, \bibinfo {author} {\bibfnamefont {S.~P.}\ \bibnamefont {Thampi}},\
  and\ \bibinfo {author} {\bibfnamefont {J.~M.}\ \bibnamefont {Yeomans}},\
  }\bibfield  {title} {\bibinfo {title} {Stabilization of active matter by
  flow-vortex lattices and defect ordering},\ }\href
  {http://dx.doi.org/10.1038/ncomms10557} {\bibfield  {journal} {\bibinfo
  {journal} {Nat. Commun.}\ }\textbf {\bibinfo {volume} {7}},\ \bibinfo {pages}
  {10557} (\bibinfo {year} {2016})}\BibitemShut {NoStop}%
\bibitem [{\citenamefont {Keber}\ \emph {et~al.}(2014)\citenamefont {Keber},
  \citenamefont {Loiseau}, \citenamefont {Sanchez}, \citenamefont {DeCamp},
  \citenamefont {Giomi}, \citenamefont {Bowick}, \citenamefont {Marchetti},
  \citenamefont {Dogic},\ and\ \citenamefont
  {Bausch}}]{KeberFC_Science345_2014}%
  \BibitemOpen
  \bibfield  {author} {\bibinfo {author} {\bibfnamefont {F.~C.}\ \bibnamefont
  {Keber}}, \bibinfo {author} {\bibfnamefont {E.}~\bibnamefont {Loiseau}},
  \bibinfo {author} {\bibfnamefont {T.}~\bibnamefont {Sanchez}}, \bibinfo
  {author} {\bibfnamefont {S.~J.}\ \bibnamefont {DeCamp}}, \bibinfo {author}
  {\bibfnamefont {L.}~\bibnamefont {Giomi}}, \bibinfo {author} {\bibfnamefont
  {M.~J.}\ \bibnamefont {Bowick}}, \bibinfo {author} {\bibfnamefont {M.~C.}\
  \bibnamefont {Marchetti}}, \bibinfo {author} {\bibfnamefont {Z.}~\bibnamefont
  {Dogic}},\ and\ \bibinfo {author} {\bibfnamefont {A.~R.}\ \bibnamefont
  {Bausch}},\ }\bibfield  {title} {\bibinfo {title} {Topology and dynamics of
  active nematic vesicles},\ }\href {http://dx.doi.org/10.1126/science.1254784}
  {\bibfield  {journal} {\bibinfo  {journal} {Science}\ }\textbf {\bibinfo
  {volume} {345}},\ \bibinfo {pages} {1135} (\bibinfo {year}
  {2014})}\BibitemShut {NoStop}%
\bibitem [{\citenamefont {Mosseri}\ and\ \citenamefont
  {Dandoloff}(2001)}]{Mosseri_2001}%
  \BibitemOpen
  \bibfield  {author} {\bibinfo {author} {\bibfnamefont {R.}~\bibnamefont
  {Mosseri}}\ and\ \bibinfo {author} {\bibfnamefont {R.}~\bibnamefont
  {Dandoloff}},\ }\bibfield  {title} {\bibinfo {title} {Geometry of entangled
  states, bloch spheres and hopf fibrations},\ }\href
  {https://doi.org/10.1088/0305-4470/34/47/324} {\bibfield  {journal} {\bibinfo
   {journal} {J. Phys. A}\ }\textbf {\bibinfo {volume} {34}},\ \bibinfo {pages}
  {10243} (\bibinfo {year} {2001})}\BibitemShut {NoStop}%
\bibitem [{\citenamefont {Chandler}\ and\ \citenamefont
  {Wolynes}(1981)}]{ChandlerD_JChemPhys74_1981}%
  \BibitemOpen
  \bibfield  {author} {\bibinfo {author} {\bibfnamefont {D.}~\bibnamefont
  {Chandler}}\ and\ \bibinfo {author} {\bibfnamefont {P.~G.}\ \bibnamefont
  {Wolynes}},\ }\bibfield  {title} {\bibinfo {title} {Exploiting the
  isomorphism between quantum theory and classical statistical mechanics of
  polyatomic fluids},\ }\href {http://dx.doi.org/10.1063/1.441588} {\bibfield
  {journal} {\bibinfo  {journal} {J. Chem. Phys.}\ }\textbf {\bibinfo {volume}
  {74}},\ \bibinfo {pages} {4078} (\bibinfo {year} {1981})}\BibitemShut
  {NoStop}%
\bibitem [{\citenamefont {Deffner}\ and\ \citenamefont
  {Zurek}(2016)}]{deffner2016foundations}%
  \BibitemOpen
  \bibfield  {author} {\bibinfo {author} {\bibfnamefont {S.}~\bibnamefont
  {Deffner}}\ and\ \bibinfo {author} {\bibfnamefont {W.~H.}\ \bibnamefont
  {Zurek}},\ }\bibfield  {title} {\bibinfo {title} {Foundations of statistical
  mechanics from symmetries of entanglement},\ }\href@noop {} {\bibfield
  {journal} {\bibinfo  {journal} {New J. Phys.}\ }\textbf {\bibinfo {volume}
  {18}},\ \bibinfo {pages} {063013} (\bibinfo {year} {2016})}\BibitemShut
  {NoStop}%
\bibitem [{\citenamefont {Sone}\ and\ \citenamefont
  {Deffner}(2021)}]{sone2021quantum}%
  \BibitemOpen
  \bibfield  {author} {\bibinfo {author} {\bibfnamefont {A.}~\bibnamefont
  {Sone}}\ and\ \bibinfo {author} {\bibfnamefont {S.}~\bibnamefont {Deffner}},\
  }\bibfield  {title} {\bibinfo {title} {Quantum and classical ergotropy from
  relative entropies},\ }\href@noop {} {\bibfield  {journal} {\bibinfo
  {journal} {arXiv preprint arXiv:2103.10850}\ } (\bibinfo {year}
  {2021})}\BibitemShut {NoStop}%
\bibitem [{\citenamefont {Machon}\ and\ \citenamefont
  {Alexander}(2014)}]{machon2014knotted}%
  \BibitemOpen
  \bibfield  {author} {\bibinfo {author} {\bibfnamefont {T.}~\bibnamefont
  {Machon}}\ and\ \bibinfo {author} {\bibfnamefont {G.~P.}\ \bibnamefont
  {Alexander}},\ }\bibfield  {title} {\bibinfo {title} {Knotted defects in
  nematic liquid crystals},\ }\href@noop {} {\bibfield  {journal} {\bibinfo
  {journal} {Phys. Rev. Lett.}\ }\textbf {\bibinfo {volume} {113}},\ \bibinfo
  {pages} {027801} (\bibinfo {year} {2014})}\BibitemShut {NoStop}%
\bibitem [{\citenamefont {Duclos}\ \emph {et~al.}(2020)\citenamefont {Duclos},
  \citenamefont {Adkins}, \citenamefont {Banerjee}, \citenamefont {Peterson},
  \citenamefont {Varghese}, \citenamefont {Kolvin}, \citenamefont {Baskaran},
  \citenamefont {Pelcovits}, \citenamefont {Powers}, \citenamefont {Baskaran}
  \emph {et~al.}}]{duclos2020topological}%
  \BibitemOpen
  \bibfield  {author} {\bibinfo {author} {\bibfnamefont {G.}~\bibnamefont
  {Duclos}}, \bibinfo {author} {\bibfnamefont {R.}~\bibnamefont {Adkins}},
  \bibinfo {author} {\bibfnamefont {D.}~\bibnamefont {Banerjee}}, \bibinfo
  {author} {\bibfnamefont {M.~S.}\ \bibnamefont {Peterson}}, \bibinfo {author}
  {\bibfnamefont {M.}~\bibnamefont {Varghese}}, \bibinfo {author}
  {\bibfnamefont {I.}~\bibnamefont {Kolvin}}, \bibinfo {author} {\bibfnamefont
  {A.}~\bibnamefont {Baskaran}}, \bibinfo {author} {\bibfnamefont {R.~A.}\
  \bibnamefont {Pelcovits}}, \bibinfo {author} {\bibfnamefont {T.~R.}\
  \bibnamefont {Powers}}, \bibinfo {author} {\bibfnamefont {A.}~\bibnamefont
  {Baskaran}}, \emph {et~al.},\ }\bibfield  {title} {\bibinfo {title}
  {Topological structure and dynamics of three-dimensional active nematics},\
  }\href@noop {} {\bibfield  {journal} {\bibinfo  {journal} {Science}\ }\textbf
  {\bibinfo {volume} {367}},\ \bibinfo {pages} {1120} (\bibinfo {year}
  {2020})}\BibitemShut {NoStop}%
\bibitem [{\citenamefont {Liu}\ \emph {et~al.}(2021)\citenamefont {Liu},
  \citenamefont {Totz}, \citenamefont {Miller}, \citenamefont {Hastewell},
  \citenamefont {Chao}, \citenamefont {Dunkel},\ and\ \citenamefont
  {Fakhri}}]{liu2021topological}%
  \BibitemOpen
  \bibfield  {author} {\bibinfo {author} {\bibfnamefont {J.}~\bibnamefont
  {Liu}}, \bibinfo {author} {\bibfnamefont {J.~F.}\ \bibnamefont {Totz}},
  \bibinfo {author} {\bibfnamefont {P.~W.}\ \bibnamefont {Miller}}, \bibinfo
  {author} {\bibfnamefont {A.~D.}\ \bibnamefont {Hastewell}}, \bibinfo {author}
  {\bibfnamefont {Y.-C.}\ \bibnamefont {Chao}}, \bibinfo {author}
  {\bibfnamefont {J.}~\bibnamefont {Dunkel}},\ and\ \bibinfo {author}
  {\bibfnamefont {N.}~\bibnamefont {Fakhri}},\ }\bibfield  {title} {\bibinfo
  {title} {Topological braiding and virtual particles on the cell membrane},\
  }\href@noop {} {\bibfield  {journal} {\bibinfo  {journal} {Proc. Natl. Acad.
  Sci. U.S.A.}\ }\textbf {\bibinfo {volume} {118}} (\bibinfo {year}
  {2021})}\BibitemShut {NoStop}%
\bibitem [{\citenamefont {Green}(1975)}]{green1975some}%
  \BibitemOpen
  \bibfield  {author} {\bibinfo {author} {\bibfnamefont {M.~L.}\ \bibnamefont
  {Green}},\ }\bibfield  {title} {\bibinfo {title} {Some {Picard} theorems for
  holomorphic maps to algebraic varieties},\ }\href@noop {} {\bibfield
  {journal} {\bibinfo  {journal} {Am. J. Math.}\ }\textbf {\bibinfo {volume}
  {97}},\ \bibinfo {pages} {43} (\bibinfo {year} {1975})}\BibitemShut {NoStop}%
\end{thebibliography}%


\begin{thebibliography}{3}%
\makeatletter
\providecommand \@ifxundefined [1]{%
 \@ifx{#1\undefined}
}%
\providecommand \@ifnum [1]{%
 \ifnum #1\expandafter \@firstoftwo
 \else \expandafter \@secondoftwo
 \fi
}%
\providecommand \@ifx [1]{%
 \ifx #1\expandafter \@firstoftwo
 \else \expandafter \@secondoftwo
 \fi
}%
\providecommand \natexlab [1]{#1}%
\providecommand \enquote  [1]{``#1''}%
\providecommand \bibnamefont  [1]{#1}%
\providecommand \bibfnamefont [1]{#1}%
\providecommand \citenamefont [1]{#1}%
\providecommand \href@noop [0]{\@secondoftwo}%
\providecommand \href [0]{\begingroup \@sanitize@url \@href}%
\providecommand \@href[1]{\@@startlink{#1}\@@href}%
\providecommand \@@href[1]{\endgroup#1\@@endlink}%
\providecommand \@sanitize@url [0]{\catcode `\\12\catcode `\$12\catcode
  `\&12\catcode `\#12\catcode `\^12\catcode `\_12\catcode `\%12\relax}%
\providecommand \@@startlink[1]{}%
\providecommand \@@endlink[0]{}%
\providecommand \url  [0]{\begingroup\@sanitize@url \@url }%
\providecommand \@url [1]{\endgroup\@href {#1}{\urlprefix }}%
\providecommand \urlprefix  [0]{URL }%
\providecommand \Eprint [0]{\href }%
\providecommand \doibase [0]{https://doi.org/}%
\providecommand \selectlanguage [0]{\@gobble}%
\providecommand \bibinfo  [0]{\@secondoftwo}%
\providecommand \bibfield  [0]{\@secondoftwo}%
\providecommand \translation [1]{[#1]}%
\providecommand \BibitemOpen [0]{}%
\providecommand \bibitemStop [0]{}%
\providecommand \bibitemNoStop [0]{.\EOS\space}%
\providecommand \EOS [0]{\spacefactor3000\relax}%
\providecommand \BibitemShut  [1]{\csname bibitem#1\endcsname}%
\let\auto@bib@innerbib\@empty
\bibitem [{\citenamefont {B{\^o}cher}(1900)}]{bocher1900theory}%
  \BibitemOpen
  \bibfield  {author} {\bibinfo {author} {\bibfnamefont {M.}~\bibnamefont
  {B{\^o}cher}},\ }\bibfield  {title} {\bibinfo {title} {The theory of linear
  dependence},\ }\href@noop {} {\bibfield  {journal} {\bibinfo  {journal} {Ann.
  Math.}\ }\textbf {\bibinfo {volume} {2}},\ \bibinfo {pages} {81} (\bibinfo
  {year} {1900})}\BibitemShut {NoStop}%
\bibitem [{\citenamefont {Cleve}\ \emph {et~al.}(1998)\citenamefont {Cleve},
  \citenamefont {Ekert}, \citenamefont {Macchiavello},\ and\ \citenamefont
  {Mosca}}]{CleveR_Proceedings454_1998}%
  \BibitemOpen
  \bibfield  {author} {\bibinfo {author} {\bibfnamefont {R.}~\bibnamefont
  {Cleve}}, \bibinfo {author} {\bibfnamefont {A.}~\bibnamefont {Ekert}},
  \bibinfo {author} {\bibfnamefont {C.}~\bibnamefont {Macchiavello}},\ and\
  \bibinfo {author} {\bibfnamefont {M.}~\bibnamefont {Mosca}},\ }\bibfield
  {title} {\bibinfo {title} {Quantum algorithms revisited},\ }\href
  {http://dx.doi.org/10.1098/rspa.1998.0164} {\bibfield  {journal} {\bibinfo
  {journal} {Proc. R. Soc. Lond. A}\ }\textbf {\bibinfo {volume} {454}},\
  \bibinfo {pages} {339} (\bibinfo {year} {1998})}\BibitemShut {NoStop}%
\bibitem [{\citenamefont {Deutsch}\ and\ \citenamefont
  {Jozsa}(1997)}]{1997DeutschJozsa}%
  \BibitemOpen
  \bibfield  {author} {\bibinfo {author} {\bibfnamefont {D.}~\bibnamefont
  {Deutsch}}\ and\ \bibinfo {author} {\bibfnamefont {R.}~\bibnamefont
  {Jozsa}},\ }\bibfield  {title} {\bibinfo {title} {Rapid solution of problems
  by quantum computation},\ }\href@noop {} {\bibfield  {journal} {\bibinfo
  {journal} {Proc. R. Soc. Lond. A}\ }\textbf {\bibinfo {volume} {439}},\
  \bibinfo {pages} {553} (\bibinfo {year} {1997})}\BibitemShut {NoStop}%
\end{thebibliography}%


\end{document}


\title{Supplementary Information:\\
Harmonic flow field representations of quantum bits and gates}

\author{Vishal P. Patil}
\address{Department of Mathematics, 
Massachusetts Institute of Technology, 
77 Massachusetts Avenue, 
Cambridge, MA 02139}
\address{School of Humanities and Sciences, 
Stanford University,
450 Serra Mall, Stanford, CA 94305}
\author{\v{Z}iga Kos}
\address{Department of Mathematics, 
Massachusetts Institute of Technology,  
77 Massachusetts Avenue, 
Cambridge, MA 02139}
\address{Faculty of Mathematics and Physics,
University of Ljubljana,
Jadranska 19, 1000 Ljubljana, Slovenia}
\author{J\"orn Dunkel}
\email{dunkel@mit.edu}
\address{Department of Mathematics, 
Massachusetts Institute of Technology, 
77 Massachusetts Avenue, 
Cambridge, MA 02139}
\date{\today}

\maketitle

\section{Defect charge representation}

Here we demonstrate the defect charge representation, another possible visualization of $n$-qubit states as flow fields. In contrast to the mapping discussed in the main text, this representation stores qubit properties entirely in the topological charge of the flow field defects. Following the general procedure described in the main text, this visualization can be constructed by choosing linear maps $\phi_j$ from the 1-qubit Hilbert space, $\mathcal{H}$, to the vector space of complex functions, $\mathcal{V}$. The map $\phi_j$ can be thought of as acting on the $j$'th qubit of an $n$-qubit state. In the defect charge representation, the single qubit basis states for the $j$'th qubit are mapped to different powers of the complex variable $z$
\begin{equation}
\phi_j\big(|0\rangle\big) = \frac{1}{z^{d^{j-1}}},\qquad
\phi_j\big(|1\rangle\big) =  z^{d^{j-1}}
\label{eq:def1}
\end{equation}
where we choose $d=3$ for our visualizations (Figs.~\ref{SI_0_to_1}-\ref{SI_defect_charge}). The expression $z^{d^j-1}$ is used as shorthand for the function $z\mapsto z^{d^j-1}$. 

\subsection*{1-qubit states}

\begin{figure*}[h]
\centering
\includegraphics[width=\textwidth]{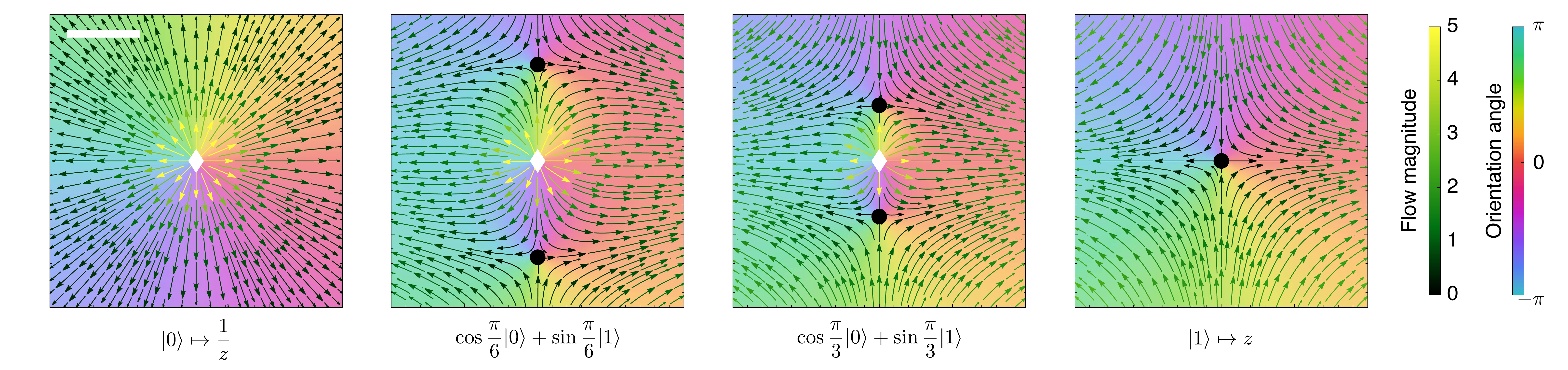}
\caption{Transition from $|0\rangle$ to $|1\rangle$. The transition is parameterized by $|\Psi\rangle=\cos\theta|0\rangle+\sin\theta|1\rangle$, where $\theta$ goes from 0 to $\pi/2$. The states are mapped to complex functions $f$ which correspond to 2D vector fields $(u,v) = (\text{Re}\,f^*, \text{Im}\,f^*)$. During the transition, two zeros (filled circles) approach the pole (white diamond) of the complex function at $z=0$. In the final state, $|\Psi'\rangle=|1\rangle$, only a zero at the origin remains. Scale bar: length~1.}
\label{SI_0_to_1}
\end{figure*}

\begin{figure*}[h]
\centering
\includegraphics[width=\textwidth]{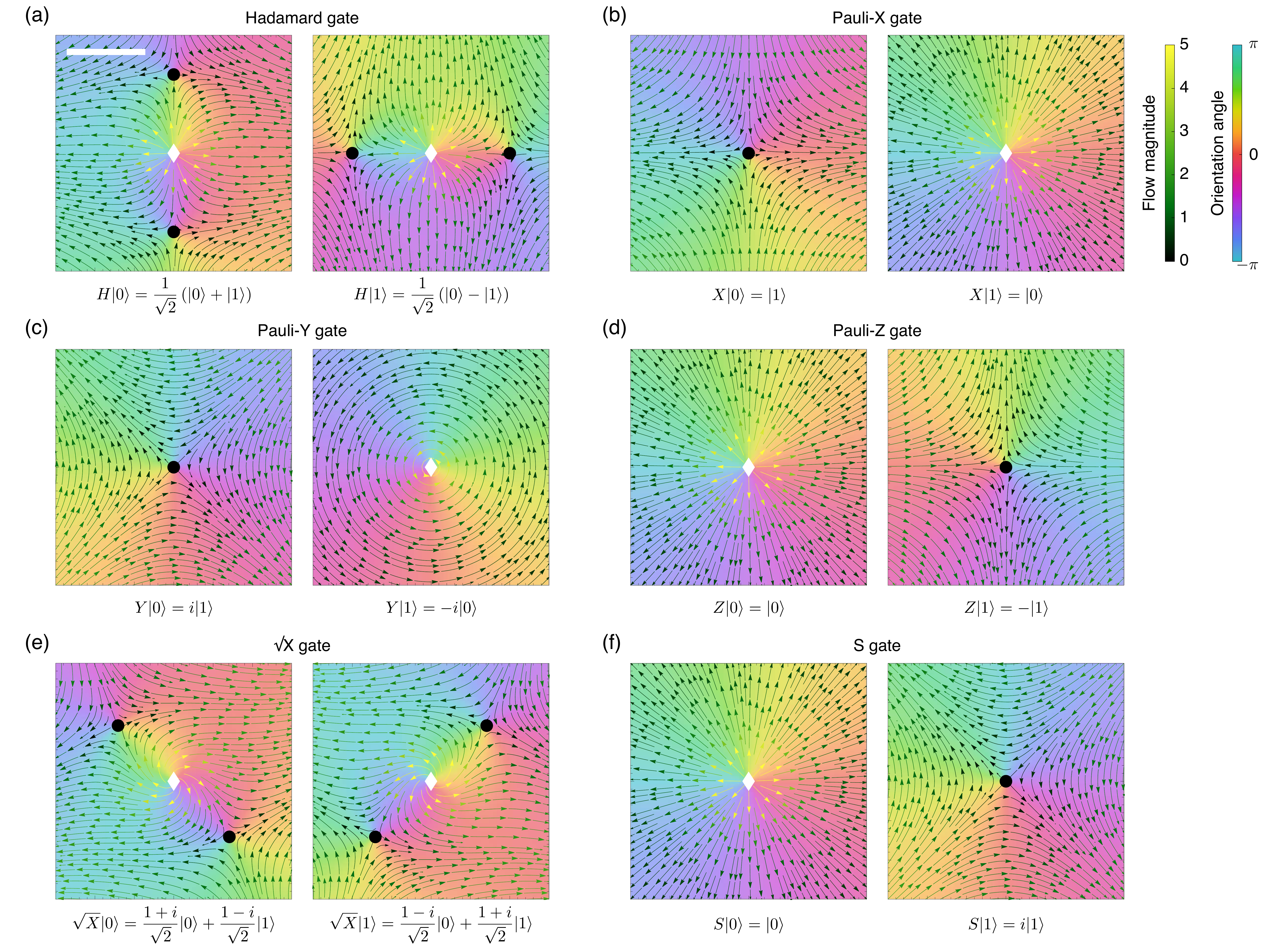}
\caption{Single qubit logic gates applied to the $|0\rangle$ and $|1\rangle$ basis states. Qubit states are mapped to complex functions $f$ with poles (white diamonds) and zeros (filled circles). (a) Hadamard gate, (b) Pauli-X gate, (c) Pauli-Y gate, (d) Pauli-Z gate, (e) $\sqrt{\text{X}}$ gate, (f) Phase shift S gate. Scale bar: length~1.}
\label{SI_single_gates}
\end{figure*}

The map $\phi_1$ is independent of $d$. For a single qubit, the defect charge representation therefore produces similar results to the representation based on defect position discussed in the main text
\begin{align*}
\phi_1\big(|0\rangle\big) = \frac{1}{z},\qquad \phi_1\big(|1\rangle\big) = z
\end{align*}
The $1$-qubit Hilbert space is mapped to the vector space of complex functions spanned by $\{1/z, z\}$. Single qubit operations, including logic gates, can be visualized in this space (Figs.~\ref{SI_0_to_1} and \ref{SI_single_gates}).

\subsection*{Multi-qubit states and innner product}

\begin{figure*}[h]
\centering
\includegraphics[width=\textwidth]{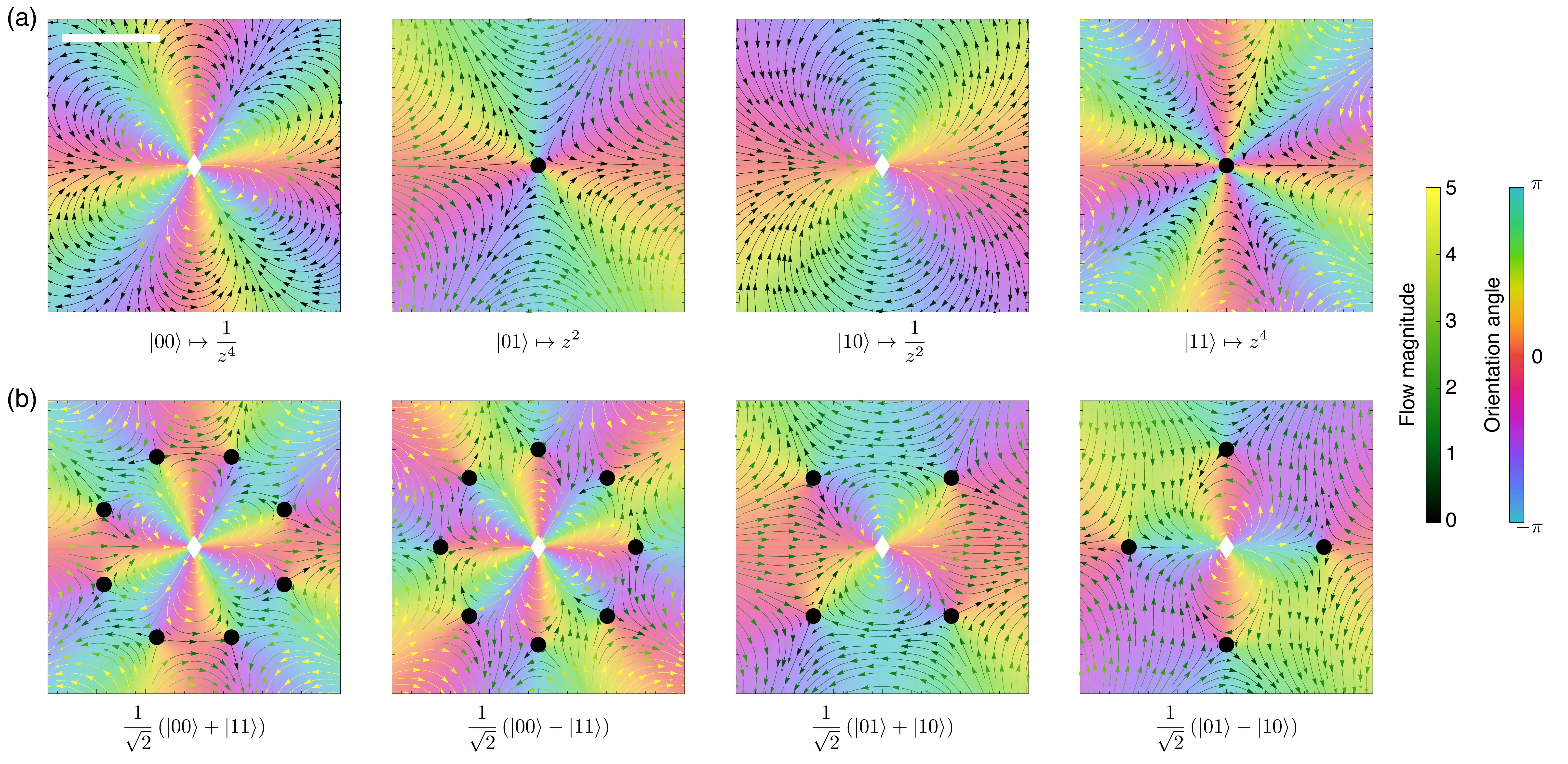}
\caption{Defect charge representation of qubit states. (a) The 2-qubit computational basis states are mapped to complex functions $f$, with either a pole (white diamond) or a zero (filled circle) at $z=0$. Different basis states are distinguished by the charge of this defect. (b) Entangled Bell states consist of a pole at $z=0$ and an additional pattern of zeros away from the pole. Scale bar: length~1.}
\label{SI_defect_charge}
\end{figure*}

Multi-qubit basis states are constructed from tensor products of single-qubit states. In the flow field picture, this is achieved by multiplying the meromorphic functions representing single-qubit states
\begin{align*}
\Phi^{(n)} \big( |\sigma_1\sigma_2...\sigma_n\rangle \big) = \prod_{j=1}^n \phi_j\big(|\sigma_j\rangle\big)
\end{align*}
In the defect charge representation, the maps $\phi_1$ and $\phi_2$ are
\begin{align*}
\phi_1\big(|0\rangle\big) =\frac{1}{z},\qquad \phi_1\big(|1\rangle\big) = z, \qquad \phi_2\big(|0\rangle\big) = \frac{1}{z^3},\qquad \phi_2\big(|1\rangle\big) = z^3
\end{align*}
which leads to the following 2-qubit flow fields
\begin{align*}
\Phi^{(2)}\big(|00\rangle\big) = \frac{1}{z^4},\quad
\Phi^{(2)}\big(|01\rangle\big) =  z^2,\quad
\Phi^{(2)}\big(|10\rangle\big) = \frac{1}{z^2},\quad
\Phi^{(2)}\big(|11\rangle\big) =  z^4
\end{align*}
The corresponding vector fields have a single defect at $z=0$, with different basis states having distinct defect charges (Fig.~\ref{SI_defect_charge}a). The full 2-qubit Hilbert space, $\mathcal{H}^{(2)} = \mathcal{H}\otimes\mathcal{H}$, is spanned by the four basis states. The entangled Bell states in $\mathcal{H}^{(2)}$ give rise to flow fields with additional defects (Fig.~\ref{SI_defect_charge}b)
\begin{align*}
\frac{1}{\sqrt{2}}\left( |00\rangle +|11\rangle  \right)\mapsto\frac{1}{\sqrt{2}}\left(\frac{1}{z^4}+z^4\right) , &\qquad
\frac{1}{\sqrt{2}}\left( |00\rangle -|11\rangle  \right)\mapsto\frac{1}{\sqrt{2}}\left(\frac{1}{z^4}-z^4\right)\\
\frac{1}{\sqrt{2}}\left( |01\rangle +|10\rangle  \right)\mapsto\frac{1}{\sqrt{2}}\left( z^2+\frac{1}{z^2}\right) , &\qquad
\frac{1}{\sqrt{2}}\left( |01\rangle -|10\rangle  \right)\mapsto\frac{1}{\sqrt{2}}\left( z^2-\frac{1}{z^2}\right)
\end{align*}

The inner product between two arbitrary states $|\Psi\rangle$ and $|\Psi'\rangle$ can be written compactly in the defect charge representation, by integrating over the unit circle $S_1$
\begin{equation*}
\langle\Psi,\Psi'\rangle=\frac{1}{2\pi}\int_0^{2\pi}\mathrm{d}\theta\,\left(\Phi^{(n)}\left[\Psi\right]\right)^* \left(\Phi^{(n)}\left[\Psi'\right]\right)
\end{equation*}

\subsection*{Linear independence}

In the defect charge representation, the general form of the maps $\phi_j$ is given by \eqref{eq:def1}
\begin{align*}
\phi_j\big(|0\rangle\big) = \frac{1}{z^{d^{j-1}}},\qquad
\phi_j\big(|1\rangle\big) =  z^{d^{j-1}}
\end{align*}
Here we show that for integers $d\geq 2$, every $n$-qubit basis state is mapped to a different power of $z$, and so the mapping $\Phi^{(n)}$ produces $2^n$ linearly independent complex functions. Consider a general $n$-qubit basis state, $|\boldsymbol{\sigma}\rangle  = |\sigma_1...\sigma_n\rangle$. This state is mapped to a complex function as follows
\begin{gather*}
\Phi^{(n)}(|\boldsymbol{\sigma}\rangle) = z^{c(\boldsymbol{\sigma})}\\
c(\boldsymbol{\sigma}) = \sum_{j=1}^n (2\sigma_j - 1)d^{j-1} = 2\sum_{j=1}^n \sigma_j d^{j-1} - \frac{d^n-1}{d-1}
\end{gather*}
Thus $c(\boldsymbol{\sigma})$ is the sum of a constant and the base $d$ representation of $\sigma_1...\sigma_n$. Base $d$ representations are unique provided $d>\sigma_j\geq 0$ for all $j$. Therefore, if $d\geq 2$, then $c(\boldsymbol{\sigma})=c(\boldsymbol{\sigma}') \Leftrightarrow  \boldsymbol{\sigma} = \boldsymbol{\sigma}'$, so every $\boldsymbol{\sigma}$ gets mapped to a different power of $z$. Different powers of $z$ are linearly independent, so this construction yields $2^n$ linearly independent states. Therefore $\ker\Phi^{(n)} = \{\mathbf{0}\}$.

\subsection*{Variable particle number}

We can also consider the question of linear independence for a more general map that acts on an arbitrary space of $k$ qubits. More concretely, let $[n] = \{1,2,...,n\}$ and let $S$ be a non-empty subset of $[n]$. Define the linear map $\Phi^{S}$ on $\mathcal{H}^{\otimes |S|}$ by its action on a basis
\begin{align*}
\Phi^{S} : \mathcal{H}^{\otimes |S|} &\rightarrow \mathcal{V}\\
\bigotimes_{j\in S} |\sigma_{j}\rangle  &\mapsto  \prod_{j\in S} \phi_{j}\left(|\sigma_{j}\rangle \right)
\end{align*}
By linearity, this defines $\Phi^S$ on the whole set $\mathcal{H}^{\otimes |S|}$. Note that the maps $\Phi^S$ are a generalization of the previously defined maps $\Phi^{(k)}:\mathcal{H}^{\otimes k} \rightarrow \mathcal{V}$
\begin{align*}
\Phi^{(k)} \big( |\sigma_{1}\sigma_1...\sigma_k\rangle \big)  =  \prod_{j = 1}^{k} \phi_{j}\left(|\sigma_{j}\rangle \right) = \prod_{j\in [k]} \phi_{j}\left(|\sigma_{j}\rangle \right)
\end{align*} 
Therefore $\Phi^{(k)} = \Phi^{[k]}$, where $[k] = \{1,2,...,k\}$. In this section, we will show that for $d\geq 3$, all the polynomials in the following set are all linearly independent
\begin{align*}
\mathcal{P} = \left\{ \Phi^{S}\left( \bigotimes_{j\in S} |\sigma_{j}\rangle \right) \; : \; S\subseteq [n], \;S\neq\emptyset, \; \boldsymbol{\sigma} \in \{0,1\}^{|S|} \right\}
\end{align*}
The condition $d\geq 3$ is necessary. For example, when $d=2$ the function $1/z$ lies in the image of both $\Phi^{(2)}$ and $\Phi^{(1)}$, so $\mathcal{P}$ is not a linearly independent set
\begin{align*}
\Phi^{(2)}\left(|1\rangle\otimes |0\rangle\right) = \phi_1(|1\rangle)\phi_2(|0\rangle) = z^{1\times 2^0 - 1\times 2^1} = z^{-1} = \Phi^{(1)}\left(|0\rangle\right)
\end{align*}
The polynomials in $\mathcal{P}$ are the image of the qubit states in the following set
\begin{align*}
\mathcal{B} = \left\{  \bigotimes_{j\in S} |\sigma_{j}\rangle  \; : \; S\subseteq [n], \;S\neq\emptyset, \; \boldsymbol{\sigma} \in \{0,1\}^{|S|}\right\}
\end{align*}
For each $k$, there are $n\choose k$ ways of selecting the set $S$, and $2^k$ values for the $k$ qubits to take. The total number of states in $\mathcal{B}$ is therefore
\begin{align*}
|\mathcal{B}| = |\mathcal{P}| = \sum_{k=1}^n {n\choose k} 2^k = (1+2)^n -1 = 3^n - 1
\end{align*}
Each state in $\mathcal{B}$ can be written down by specifying whether the $j$'th qubit is $0, 1$ or not included in the state. The states in $\mathcal{B}$ can therefore be written in terms of ternary strings, $\boldsymbol{\tau} = \tau_1\tau_2...\tau_n$ where each digit is $0,1$ or $2$. Define a subset $S(\boldsymbol{\tau})$ of $[n]$ as follows
\begin{align*}
S(\boldsymbol{\tau}) = \{j \; :\; \tau_j \neq 1 \}
\end{align*}
Since we require $S$ to be a non-empty, the string $\boldsymbol{\tau} = \mathbf{1} = (1,1,...,1)$ must be excluded. With this notation $\mathcal{B}$ becomes
\begin{align*}
\mathcal{B} = \left\{  \bigotimes_{j\in S(\boldsymbol{\tau})} |\tau_{j}/2\rangle \; : \;  \boldsymbol{\tau} \in \{0,1,2\}^n\backslash \{\mathbf{1}\}  \right\}
\end{align*}
In this form, we clearly have $|\mathcal{B}| = 3^n - 1$. Similarly, $\mathcal{P}$ is
\begin{align*}
\mathcal{P} = \left\{ \Phi^{S(\boldsymbol{\tau})} \left( \bigotimes_{j\in S(\boldsymbol{\tau})} |\tau_{j}/2\rangle\right) \; : \;  \boldsymbol{\tau} \in \{0,1,2\}^n\backslash \{\mathbf{1}\}  \right\}
\end{align*}
As before, each element of $\mathcal{P}$ is a power of $z$.
\begin{gather*}
\Phi^{S(\boldsymbol{\tau}) }\left( \bigotimes_{j\in S(\boldsymbol{\tau})} |\tau_{j}/2\rangle \right) = z^{c(\boldsymbol{\tau})}\\
c(\boldsymbol{\tau}) = \sum_{j=1}^n (\tau_j - 1)d^{j-1} = \sum_{j=1}^n \tau_j d^{j-1} - \frac{d^n-1}{d-1}
\end{gather*}
The exponent $c(\boldsymbol{\tau})$ is the sum of a constant and the base $d$ representation of $\tau_1...\tau_n$. Base $d$ representations are unique provided $d>\sigma_j\geq 0$ for all $j$. Therefore, if $d\geq 3$, every $\boldsymbol{\tau}$ gets mapped to a different power of $z$. Different powers of $z$ are linearly independent, so this construction shows that $\mathcal{P}$ is linearly independent.

\section{Defect position representation}

In this section we analyze properties of the visualization presented in the main text, the defect position representation. This visualization maps $n$-qubit states to vector fields using translates of the functions $q_0(z) = z^{-d}$ and $q_1(z) = z^d$ 
\begin{align*}
|\boldsymbol{\sigma}\rangle  = |\sigma_1...\sigma_n\rangle \mapsto (z-a_1)^{(2\sigma_1-1) d}...(z-a_n)^{(2\sigma_n-1) d} = q_{\boldsymbol{\sigma}}(z) = h_{\boldsymbol{\sigma}}(z)^d
\end{align*}
where the $a_j$'s are distinct but can otherwise be chosen arbitrarily. When $d=1$ and $a_1=0$, the single qubit vector fields coincide with those of defect charge representation (Figs.~\ref{SI_0_to_1} and \ref{SI_single_gates}).

\subsection*{Linear independence}

In this representation, the computational basis states are mapped to rational functions, $h_{\boldsymbol{\sigma}}(z)^d$.
To show that the $2^n$ rational functions $\{h_{\boldsymbol{\sigma}}^d\}$ are linearly independent, we use an argument due to Bjorn Poonen.

\begin{thm}
Let $f_1,...,f_N, g_1,...,g_N$ be nonzero rational functions from $\mathbb{C} \rightarrow \mathbb{C}$ such that no two $f_i$ are linearly dependent. Define the degree, $\deg f$, of a rational function $f$, by the maximum of the degrees of the numerator and denominator of $f$ when written as a fraction in its lowest terms. Suppose $\deg f_i, \deg g_i \leq M $ for all $i$, where $M\in \mathbb{Z}_{\geq 0}$. Then for any $d> 14^N M$, the expression $\sum_{i=1}^N g_i f_i^d$ is not identically 0.
\end{thm}

This theorem implies that the functions $\{h_{\boldsymbol{\sigma}}^d\}$ are linearly independent. In particular, if the family $\{h_{\boldsymbol{\sigma}}^d\}$ were linearly dependent, then we could find constants $\lambda_{\boldsymbol{\sigma}} \in \mathbb{C}$ which are not all 0, such that 
\begin{align*}
\sum_{\boldsymbol{\sigma}} \lambda_{\boldsymbol{\sigma}}h_{\boldsymbol{\sigma}}^d \equiv 0
\end{align*}
Let $S$ be the set of strings $\boldsymbol{\sigma}$ for which $\lambda_{\boldsymbol{\sigma}}$ is nonzero. Then
\begin{align*}
\sum_{\boldsymbol{\sigma}\in S} \lambda_{\boldsymbol{\sigma}}h_{\boldsymbol{\sigma}}^d \equiv 0
\end{align*}
Since $\lambda_{\boldsymbol{\sigma}}, h_{\boldsymbol{\sigma}}$ are nonzero rational functions, and by construction, no two $h_{\boldsymbol{\sigma}}$ are linearly dependent, this violates the above theorem. In addition, observe that there are $2^n$ functions $h_{\boldsymbol{\sigma}}$, and $\deg h_{\boldsymbol{\sigma}} \leq n$. Therefore, the functions $\{h_{\boldsymbol{\sigma}}^d\}$ are linearly independent for all $d > 14^{2^n} n$. This represents an upper bound on the smallest $d$ needed for linear independence. In other words, there is a $d\leq 14^{2^n} n +1$ for which the functions $\{h_{\boldsymbol{\sigma}}^d\}$ are linearly independent.

\bigskip

\noindent 
\textbf{Proof:} 
We induct on the number of functions $N$. Concretely, the induction hypothesis states that for all nonzero rational functions $f_i, g_i$ with no two $f_i$ linearly dependent, the following is true
\begin{align*}
\sum_{i=1}^{N-1} g_i f_i^d \neq 0 , \quad \forall d>14^{N-1} \max \{\deg f_i, \deg g_i\}    
\end{align*} 
The base case is clear, since $g_1f_1^d$ is nonzero for any $d$. Suppose now that we have some $f_i$ and $g_i$ satisfying the conditions of the theorem such that $\sum_{i=1}^N g_i f_i^d = 0$, for some $d>14^N M$, where $\deg f_i, \deg g_i \leq M $ for all $i$. Divide by $g_N f_N^d$ and let $\tilde{g}_i = g_i/g_N$, $\tilde{f}_i = f_i/f_N$
\begin{align*}
\sum_{i=1}^{N-1} \tilde{g}_i \tilde{f}_i^d + 1 = 0
\end{align*}
where $\deg \tilde{f}_i, \deg \tilde{g}_i \leq 2M $. Since no two $f_i$ are linearly dependent, each $\tilde{f}_i$ is nonconstant. Differentiating the above sum gives
\begin{align*}
\sum_{i=1}^{N-1} \left( \frac{\tilde{g}'_i}{\tilde{g}_i} + d \frac{\tilde{f}'_i}{\tilde{f}_i}   \right)\tilde{g}_i \tilde{f}_i^d = \sum_{i=1}^{N-1} G_i \tilde{f}_i^d = 0
\end{align*}
where
\begin{align*}
G_i = \left( \frac{\tilde{g}'_i}{\tilde{g}_i} + d \frac{\tilde{f}'_i}{\tilde{f}_i}   \right)\tilde{g}_i
\end{align*}
By construction, the $\tilde{f}_i$ are nonzero with $\deg \tilde{f}_i \leq 2M$, and no two are linearly dependent. To use the induction hypothesis, we therefore need to bound $\deg G_i$ and show that the $G_i$ are nonzero for large $d$. Starting with the latter, observe that
\begin{align*}
\frac{G_i}{\tilde{g}_i} = \frac{d}{dz} \log\left( \tilde{g}_i \tilde{f}_i^d\right)
\end{align*}
$G_i$ is therefore nonzero whenever $\tilde{g}_i \tilde{f}_i^d$ is nonconstant. Since $\deg \tilde{g}_i \leq 2M$, and $\tilde{f}_i$ is nonconstant, we have that $\tilde{g}_i \tilde{f}_i^d$ is nonconstant for all $d>2M$. Since we are assuming $d>14^N M$ and $N\geq 1$, we therefore have that $G_i$ is nonzero. Now consider $\deg G_i$.
\begin{align*}
\deg \tilde{g}'_i \leq 4M , \quad \deg \left(\frac{\tilde{g}'_i}{\tilde{g}_i} \right) \leq 6M, \quad \deg\left( \frac{\tilde{g}'_i}{\tilde{g}_i} + d \frac{\tilde{f}'_i}{\tilde{f}_i}   \right) \leq 12M, \quad \deg G_i \leq 14M
\end{align*}
Finally, observe that $14^N M \geq 14^{N-1} \max \{\deg \tilde{f}_i, \deg G_i\}$. By the induction hypothesis, we therefore have that $\sum_{i=1}^{N-1} G_i \tilde{f}_i^d$ is nonzero for all $d>14^N M$, which is a contradiction.

\bigskip

\subsection*{Lower bound on $d$}

A lower bound on the smallest $d$ for which the functions $\{h_{\boldsymbol{\sigma}}^d\}$ are linearly independent follows from obtaining polynomials from the rational functions $h_{\boldsymbol{\sigma}}^d$. In particular, we can clear the poles of $h^d_{\boldsymbol{\sigma}}$ through multiplication with $(z-a_1)^d...(z-a_n)^d$
\begin{align*}
p_{\boldsymbol{\sigma}}(z)  = (z-a_1)^d...(z-a_n)^d h^d_{\boldsymbol{\sigma}} (z)= (z-a_1)^{2\sigma_1 d}...(z-a_n)^{2\sigma_n d}
\end{align*}
The functions $\{p_{\boldsymbol{\sigma}}\}$ are linearly independent if and only if the $\{h^d_{\boldsymbol{\sigma}}\}$ are linearly independent, so it suffices to focus on the $\{p_{\boldsymbol{\sigma}}\}$. Observe that there are $n\choose k$ states $|\boldsymbol{\sigma}\rangle$ with exactly $k$ ones and $n-k$ zeros, so there are $n\choose k$ polynomials $p_{\boldsymbol{\sigma}}$ with degree $2kd$. There are therefore $\sum_{j\leq k}{n \choose j}$ polynomials $p_{\boldsymbol{\sigma}}$ with degree at most $2kd$, and these must all be linearly independent. Polynomials of degree at most $2kd$ span a $(2kd+1)$-dimensional vector space, so for linear independence, we need
\begin{align*}
2kd +1  \geq \sum_{j=0}^k {n\choose j} , \quad \forall\; 0\leq k\leq n
\end{align*}
For odd $n$, the sum of the first $\lfloor n/2 \rfloor$ binomial coefficients is $2^{n-1}$, which gives the following bound
\begin{align*}
(n-1)d +1 \geq 2^{n-1}
\end{align*}
Thus $d$ must grow at least as fast as $d \sim n^{-1} 2^n$.

\subsection*{Linear independence of 2-qubit representation}

In the main text we present a 2-qubit construction with $d=1$
\begin{align*}
|\boldsymbol{\sigma}\rangle \mapsto h_{\boldsymbol{\sigma}}(z) = (z+1)^{2\sigma_1 -1}(z-1)^{2\sigma_2 -1}
\end{align*}
We can show that this construction is valid and produces 4 linearly independent states. To show this, it suffices to find $b_1,b_2,b_3,b_4 \in \mathbb{C}$ such that the vectors, $\left(h_{\boldsymbol{\sigma}}(b_1),h_{\boldsymbol{\sigma}}(b_2),h_{\boldsymbol{\sigma}}(b_3),h_{\boldsymbol{\sigma}}(b_4)\right), \; \boldsymbol{\sigma}\in\{0,1\}^2$, are linearly independent. Equivalently, the matrix of these vectors, which we denote $M$, must be nonsingular
\begin{align*}
M = 
\begin{pmatrix}
h_{00}(b_1) && h_{01}(b_1) && h_{10}(b_1) && h_{11}(b_1)\\
h_{00}(b_2) && h_{01}(b_2) && h_{10}(b_2) && h_{11}(b_2)\\
h_{00}(b_3) && h_{01}(b_3) && h_{10}(b_3) && h_{11}(b_3)\\
h_{00}(b_4) && h_{01}(b_4) && h_{10}(b_4) && h_{11}(b_4)
\end{pmatrix}
\end{align*}
Choosing $(b_1, b_2, b_3, b_4) = (0,i,2i,3i)$ yields a nonsingular $M$. Note that the order of the $b_i$'s does not affect the invertibility of $M$.

\subsection*{Linear independence of 4-qubit representation}

In the main text we present a 4-qubit construction with $d=3$
\begin{align*}
|\boldsymbol{\sigma}\rangle \mapsto h_{\boldsymbol{\sigma}}(z)^3 = \left[(z+1)^{2\sigma_1 -1}(z-i)^{2\sigma_2 -1}(z-1)^{2\sigma_3 -1}(z+i)^{2\sigma_4 -1}\right]^3
\end{align*}
As above, to show that this construction is valid and produces 16 linearly independent states, it suffices to find $b_1,...b_{16} \in \mathbb{C}$ such that the vectors, $(h_{\boldsymbol{\sigma}}(b_1)^3,...,h_{\boldsymbol{\sigma}}(b_{16})^3), \; \boldsymbol{\sigma}\in\{0,1\}^4$, are linearly independent. Equivalently, the matrix of these vectors, $M$, must be nonsingular
\begin{align*}
M = 
\begin{pmatrix}
h_{0000}(b_1)^3 && h_{0001}(b_1)^3 && \hdots && h_{1111}(b_1)^3\\
h_{0000}(b_2)^3 && h_{0001}(b_2)^3 &&\hdots && h_{1111}(b_2)^3\\
\vdots  && \vdots && \ddots && \vdots\\
h_{0000}(b_{16})^3 && h_{0001}(b_{16})^3 && \hdots && h_{1111}(b_{16})^3
\end{pmatrix}
\end{align*}
Choosing the $b_i$'s to be the elements of the following set yields a nonsingular $M$
\begin{align*}
\left\{\left(x+\frac{1}{2}\right)+i\left(y-\frac{1}{2}\right) : x,y = 0,1,2,3 \right\}
\end{align*}
In addition, this result implies that linear independence also holds for the following 3-qubit construction contained within the above 4-qubit construction 
\begin{align*}
|\boldsymbol{\sigma}\rangle \mapsto h_{\boldsymbol{\sigma}}(z)^3 = \left[(z+1)^{2\sigma_1 -1}(z-i)^{2\sigma_2 -1}(z-1)^{2\sigma_3 -1}\right]^3
\end{align*}

\subsection*{Inner Product}

Here we show that the inner product constructed for the defect position representation is a valid inner product on $\Phi^{(n)}\left(\mathcal{H}^{(n)}\right)$. The construction of the inner product uses the following mapping
\begin{align*}
\pi: \Phi^{(n)}\left(\mathcal{H}^{(n)}\right) \rightarrow \mathbb{C}^{2^n} , \qquad f \mapsto \boldsymbol{v} = \left[ f(\alpha), Df(\alpha), D^2f(\alpha),..., D^{2^n-1}f(\alpha)  \right]
\end{align*}
where $\alpha \in \mathbb{C}$ and $D$ is the derivative operator. The matrix $B$ is defined to be the $2^n\times 2^n$ matrix whose columns are the vectors $\pi\circ\Phi^{(n)}\left(|\boldsymbol{\sigma}\rangle\right)$, where $\boldsymbol{\sigma}\in\{0,1\}^n$. Provided $B$ is invertible, we set $P = (B^{-1})^\dagger B^{-1}$ and define the inner product on $\Phi^{(n)}\left(\mathcal{H}^{(n)}\right)$ by
\begin{align*}
\langle f_1, f_2\rangle = \pi(f_1)^{\dagger}\, P\, \pi(f_2)
\end{align*}
By construction, the $2^n$ functions $\{\Phi^{(n)}\left(|\boldsymbol{\sigma}\rangle\right) : \boldsymbol{\sigma} \in \{0,1\}^n \}$, are orthonormal with respect to this inner product. This follows from the fact that these functions are sent to the column vectors of $B$ under the mapping $\pi$. It remains to show that $B$ is invertible for some choice of $\alpha \in \mathbb{C}$. Using $q_{\boldsymbol{\sigma}}(z) = h_{\boldsymbol{\sigma}}(z)^d = \Phi^{(n)}\left(|\boldsymbol{\sigma}\rangle\right)$, we can write down $B$
\begin{align*}
B = B(\alpha)= \left[\pi\left(q_{0...0}\right),...,\pi\left(q_{1...1}\right) \right]=
\begin{pmatrix*}[r]
q_{0...0}(\alpha) & \hdots & q_{1...1}(\alpha)\\
Dq_{0...0}(\alpha)  & \hdots & Dq_{1...1}(\alpha)\\
\vdots   & \ddots & \vdots\\
D^{2^n-1}q_{0...0}(\alpha)  & \hdots & D^{2^n-1}q_{1...1}(\alpha)
\end{pmatrix*}
\end{align*}
Here, the $\alpha$ dependence of $B$ has been made explicit. The determinant, $\det B(\alpha)$, is the Wronskian of the functions $\{q_{\boldsymbol{\sigma}}\}$. The Wronskian of a set of analytic functions is identically zero if and only if the functions are linearly dependent~\cite{bocher1900theory}. Since the $\{q_{\boldsymbol{\sigma}}\}$ are linearly independent (as shown above), there must therefore be some $\alpha \in \mathbb{C}$ for which $B$ is invertible.

\subsection*{Separable states and factorizable states}

Here we show that equiangular defect halos in the defect position visualization are a necessary and sufficient condition for separability of the underlying qubit state. More prescisely, a qubit state is separable if and only if the corresponding flow field consists of basis defects each surrounded by an equiangular (i.e. regular polygon) halo of defects with no defects left over. We work with a visualization with charge $d$, and with basis defects at $a_1,...,a_n \in \mathbb{C}$. Consider a general separable state and its corresponding flow field
\begin{align*}
    |\boldsymbol{\sigma}\rangle &= \bigotimes_{j=1}^n \left( \alpha_j |0 \rangle + \beta_j |1\rangle\right)\\
    \Phi^{(n)}\big( |\boldsymbol{\sigma}\rangle \big) &= \prod_{j=1}^n \left(z - a_j\right)^{-d} \left(\alpha_j + \beta_j (z-a_j)^{2d} \right)
\end{align*}
where $\alpha_j,\beta_j\in \mathbb{C}$. Let $\gamma_{j,1},...,\gamma_{j,2d}$ be the solutions of $\alpha_j + \beta_j (z-a_j)^{2d}=0$
\begin{align*}
    \gamma_{j,k} = a_j + \left| \frac{\alpha_j}{\beta_j}\right|^{\frac{1}{2d}} \exp\left( \frac{i}{2d}\arg \left(\frac{-\alpha_j}{\beta_j}\right) + \frac{k\pi i}{d}\right)
\end{align*}
Consider first the case where $\alpha_j,\beta_j\neq 0$ and the $(2d+1)n$ numbers, $\{a_j, \gamma_{j,k}\}$, are all distinct. Then the complex function $\Phi^{(n)}\big( |\boldsymbol{\sigma}\rangle \big)$ will have poles of degree $d$ at each $a_j$, surrounded by an equiangular halo of zeros of degree 1 at each $\gamma_{j,k}$, and a pole of degree $nd$ at infinity. If some of the numbers $\{a_j, \gamma_{j,k}\}$ coincide, or if $\gamma_{j,k}=\infty$ (as is the case when $\beta_j = 0$), then the degree of corresponding pole or zero must be adjusted accordingly. However, the overall configuration of defects can still be interpreted as poles surrounded by equiangular halos of zeros.

Now suppose we have a complex function $f$ which defines the flow field for some $n$-qubit state $|\boldsymbol{\sigma}\rangle$, so $f = \Phi^{(n)}\big( \boldsymbol{\sigma}\big)$. Suppose that $f$ has $n$ poles of degree $d$ at points $a_j$, surrounded by halos of $2d$ zeros with degree 1 at points $\gamma_{j,k}$. Some of the numbers $\{a_j, \gamma_{j,k}\}$ may coincide, however the $a_j$'s must be distinct, as is requried by the defect position representation. Furthermore, the $\gamma_{j,k}$'s are allowed to lie at infinity. Observe that if $\gamma_{j,k} = \infty$ for some $k$ then $\gamma_{j,k}=\infty$ for all $k$. Define $\alpha'_j,\beta'_j \in \mathbb{C}$ as follows
\begin{align*}
    \alpha'_j = \begin{cases}
    1 \; &\text{if } \gamma_{j,k} = \infty\\
    (\gamma_{j,k} - a_j)^{2d} \; &\text{otherwise}\end{cases} \qquad \qquad
    \beta'_j = \begin{cases}
    0 \; &\text{if } \gamma_{j,k} = \infty\\
    1 \; &\text{otherwise}\end{cases} 
\end{align*}
Since the $\gamma_{j,k}$'s are assumed to form regular $2d$-gons centered at the $a_j$'s, the expressions for $\alpha'_j,\beta'_j$ do not depend on $k$. Consider the separable state
\begin{align*}
    |\boldsymbol{\tau}\rangle  = \bigotimes_{j=1}^n \left( \alpha'_j |0 \rangle + \beta'_j |1\rangle\right)
\end{align*}
By construction, we have
\begin{align*}
    \Phi^{(n)}\big( |\boldsymbol{\tau}\rangle \big) = \lambda f= \lambda \Phi^{(n)}\big( |\boldsymbol{\sigma}\rangle \big)
\end{align*}
for some $\lambda \in \mathbb{C}$. Since $\Phi^{(n)}$ is linear, we have $|\boldsymbol{\tau}\rangle = \lambda|\boldsymbol{\sigma}\rangle$ so $|\boldsymbol{\sigma}\rangle$ is separable.

Finally consider $n$-qubit states which are factorziable into the product of a 1-qubit state and an $(n-1)$-qubit state. Equiangular defect halos in the flow field picture are a necessary condition for this factorization to exist, as follows from a similar argument as above. However, the existence of these defect halos is not a sufficient condition for such a factorization. To see this, observe that equiangular defect halos around a basis defect $a_j$ are allowed to collapse onto $a_j$ or be pushed to infinity. Therefore, a pole at $a_j$ can be considered to represent a basis defect together with an equiangular halo of defects. However, entangled states which do not permit any factorizations typically have poles at the basis defect locations (Main text fig.~3).

\subsection*{Deutsch-Josza algorithm}

\begin{figure*}[t]
	\centering
	\includegraphics[width=\textwidth]{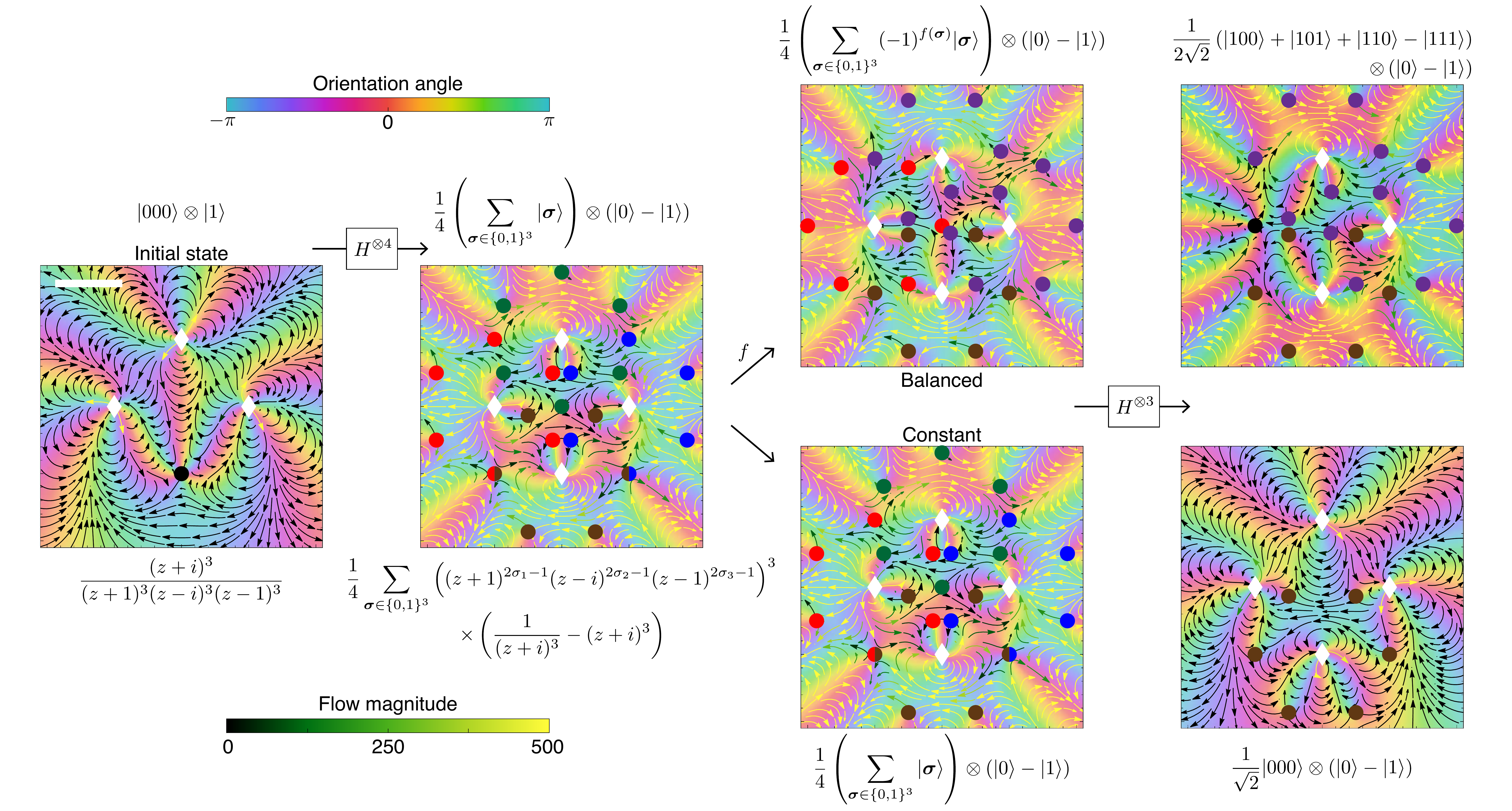}
	\caption{Flow field visualization of the  Deutsch-Josza algorithm.
		The Deutsch-Josza algorithm can be visualized on 3 qubits, at $-1, +i, +1$, with an ancillary qubit at $-i$. The algorithm involves applications of the single qubit Hadamard gate, $H$ (Fig.~\ref{SI_single_gates}) and a function which is either balanced or constant~\cite{CleveR_Proceedings454_1998}. In this example, the balanced function satisfies $f(000) = f(001) = f(010) = f(111)  = 0, \; f(011) = f(100) = f(101) = f(110) = 1$ . The flow fields are characterized by their poles (white diamonds) and zeros (filled circles). The algorithm passes through intermediate entangled states, producing flow fields with zeros that do not form equiangular halos (purple circles). The presence of equiangular halo zeros (red, green, blue and brown circles) in certain states indicates separability or factorizability. Scale bar: length~1.} 
	\label{SI_DJ_fig}
\end{figure*}

In addition to the algorithms demonstrated in the main text, the defect position representation can also be used to visualize the Deutsch-Josza algorithm~\cite{1997DeutschJozsa} on 3 qubits and 1 ancillary qubit (Fig.~\ref{SI_DJ_fig}). The defect halo criterion allows for the visual identification of entanglement. In particular, the 3-bit balanced function $f$ that is implemented during the Deutsch-Josza algorithm in Fig.~\ref{SI_DJ_fig} has been chosen to produce entanglement. This function is 
\begin{align*}
	f(000) = f(001) = f(010) = f(111)  = 0, \qquad f(011) = f(100) = f(101) = f(110) = 1
\end{align*}
Ignoring the ancillary qubit and the overall normalization of the state, implementing $f$ produces the following state (Fig.~\ref{SI_DJ_fig}, third column, top row)
\begin{align*}
	\sum_{\boldsymbol{\sigma}\in\{0,1\}^3}(-1)^{f(\boldsymbol{\sigma})}|\boldsymbol{\sigma}\rangle &=  |000\rangle + |001\rangle + |010\rangle - |011\rangle - |100\rangle - |101\rangle - |110\rangle + |111\rangle \\
	&= |0\rangle \otimes \left(|00\rangle + |01\rangle + |10\rangle - |11\rangle\right) + |1\rangle \otimes \left( - |00\rangle - |01\rangle - |10\rangle + |11\rangle \right) \\
	&= \left( |0\rangle - |1\rangle\right) \otimes \left(|00\rangle + |01\rangle + |10\rangle - |11\rangle\right) \\
	&= \left( |0\rangle - |1\rangle\right) \otimes \big[ |0\rangle \otimes \left( |0\rangle + |1\rangle \right) + |1\rangle \otimes \left( |0\rangle - |1\rangle \right) \big]
\end{align*}
The first qubit factorizes out, but the second and third qubits do not factorize. Thus the state has entanglement, but the corresponding flow field will have a defect halo around the first basis defect (Fig.~\ref{SI_DJ_fig}, third column, top row).

\section{Stereographic projection}

\begin{figure*}[h]
\centering
\includegraphics[width=\textwidth]{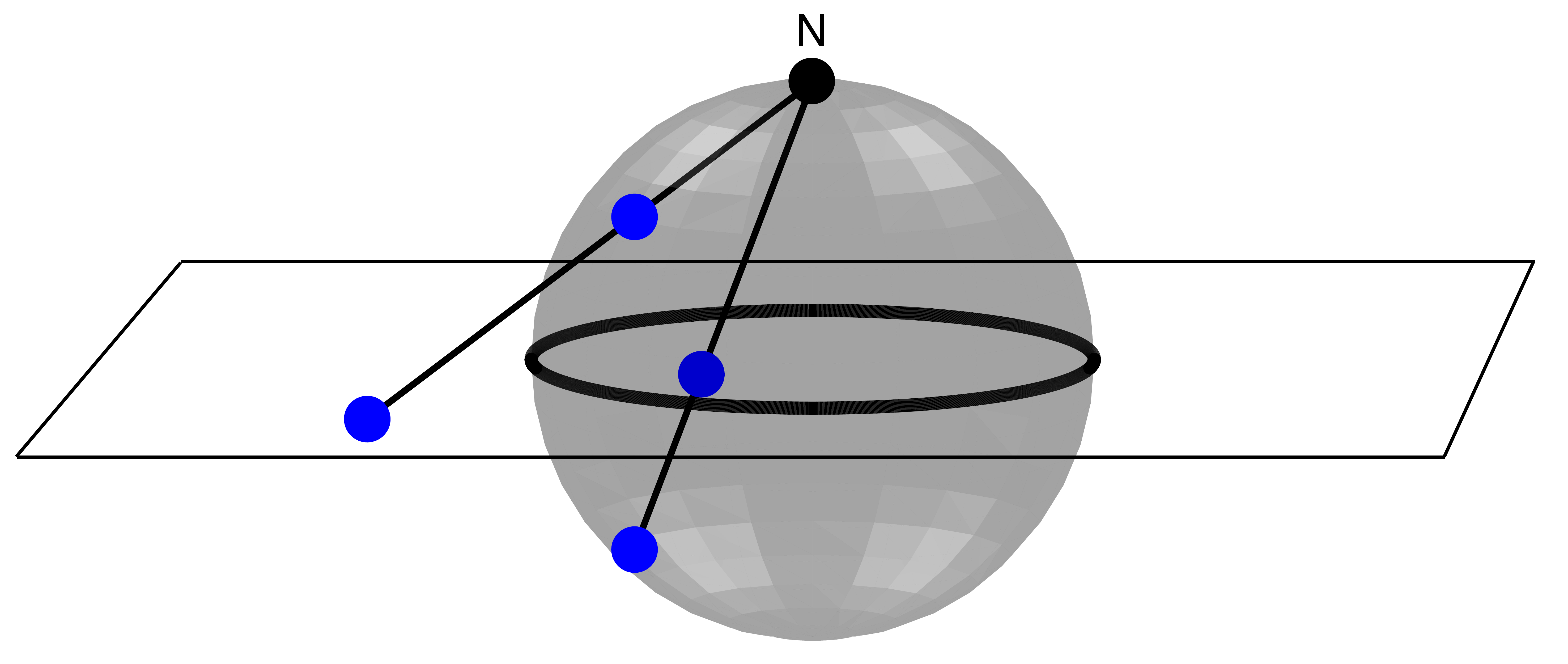}
\caption{Stereographic projection maps the sphere, $S^2$, to the plane, $\mathbb{R}^2$.}
\label{SI_stereographic}
\end{figure*}

\begin{figure*}[h]
\centering
\includegraphics[width=\textwidth]{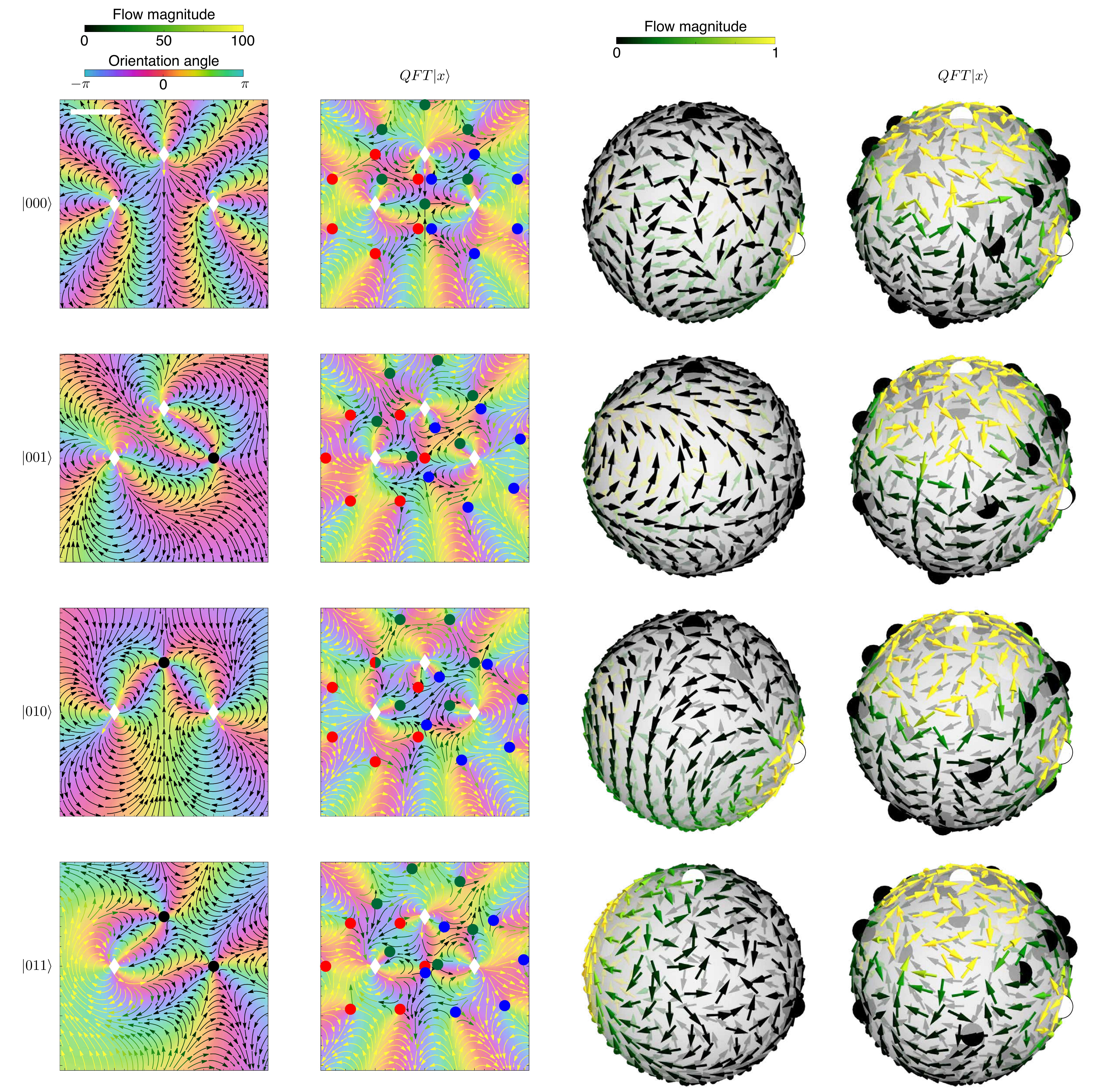}
\caption{Visualization of the 3-qubit quantum Fourier transform on $|000\rangle, |001\rangle, |010\rangle, |011\rangle$, using the defect position representation. Qubit states are mapped to complex functions with poles (white diamonds) and zeros (filled circles) (columns 1 and 2). These can in turn be mapped to the sphere (columns 3 and 4), where poles and zeros are denoted by white and black circles respectively. The vector fields for the computational basis states (columns 1 and 3) have poles and zeros at the points $-1,i,+1$. The quantum Fourier transform maps computational basis states to separable states (columns 2 and 4). In the plane, separable states are characterized by equiangular defect halos centered at the locations of the basis defects (column 2; red, green and blue circles). Scale bar: length~1.}
\label{SI_qft1}
\end{figure*}

\begin{figure*}[h]
\centering
\includegraphics[width=\textwidth]{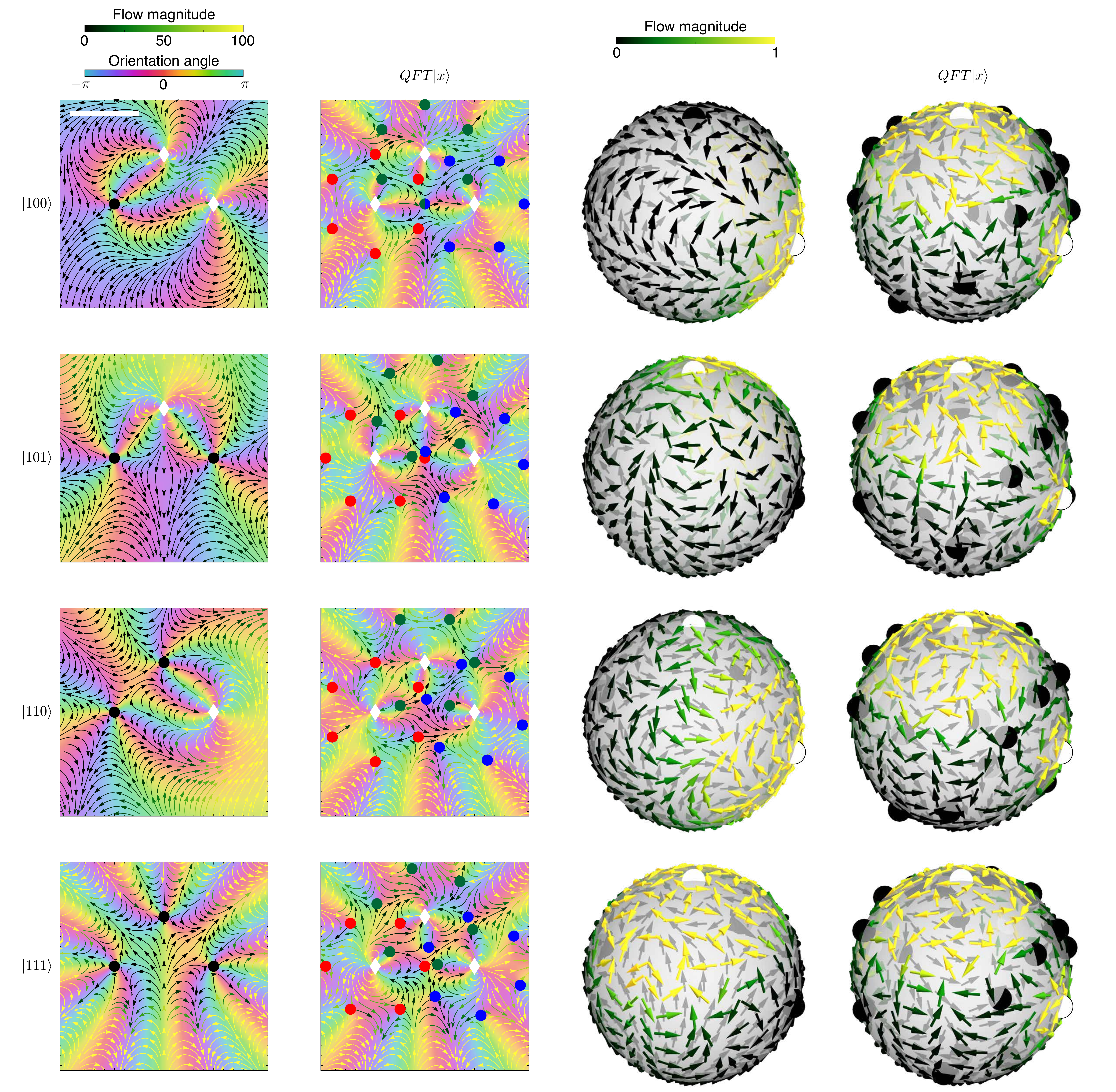}
\caption{Visualization of the 3-qubit quantum Fourier transform on $|100\rangle, |101\rangle, |110\rangle, |111\rangle$, using the defect position representation. Qubit states are mapped to complex functions with poles (white diamonds) and zeros (filled circles) (columns 1 and 2). These can in turn be mapped to the sphere (columns 3 and 4), where poles and zeros are denoted by white and black circles respectively. The vector fields for the computational basis states (columns 1 and 3) have poles and zeros at the points $-1,i,+1$. The quantum Fourier transform maps computational basis states to separable states (columns 2 and 4). In the plane, separable states are characterized by equiangular defect halos centered at the locations of the basis defects (column 2; red, green and blue circles). Scale bar: length~1.}
\label{SI_qft2}
\end{figure*}

The planar vector fields we study here can be mapped to vector fields on a sphere through inverse stereographic projection (Figs.~\ref{SI_stereographic}-\ref{SI_qft2}). Treating the sphere $S^2$ as embedded in $\mathbb{R}^3$, the stereographic projection map and its inverse are
\begin{align*}
g(x,y,z) &= (X,Y) = \left(\frac{x}{1-z}, \frac{y}{1-z}\right)\\
g^{-1}(X,Y) &= (x,y,z) =  \left(\frac{2X}{1+X^2+Y^2}, \frac{2Y}{1+X^2+Y^2}, \frac{-1+X^2+Y^2}{1+X^2+Y^2}\right)
\end{align*}
where $x^2+y^2+z^2=1$. The derivative of the inverse map, $Dg^{-1}$, maps vector fields on the plane to vector fields on the sphere. In particular, a vector field $V=(V_1,V_2)$ on $\mathbb{R}^2$ is mapped to a vector field $U$ on $S^2$ by 
\begin{align*}
U(p) = Dg^{-1}|_{g(p)} V(g(p))
\end{align*}
where $p$ is a point on $S^2$. In coordinates this becomes
\begin{align*}
\begin{pmatrix}
U_1(x,y,z) \\
U_2(x,y,z) \\
U_3(x,y,z)
\end{pmatrix} = 
\begin{pmatrix}
-x^2 + 1-z && -xy\\
-xy && -y^2 +1-z\\
x(1-z) && y(1-z)
\end{pmatrix}
\begin{pmatrix}
V_1\left(\frac{x}{1-z},\frac{y}{1-z} \right) \\
V_2\left(\frac{x}{1-z},\frac{y}{1-z} \right)
\end{pmatrix}
\end{align*}
Away from the north pole, defined by $z=1$, the singularities of $U$ can be mapped to singularities of $V$ and vice versa, using the mappings $g$ and $g^{-1}$. To understand the behavior of the vector field at the north pole, it is convenient to use spherical polar coordinates $(\theta,\phi)$
\begin{align*}
x = \cos\phi \sin\theta , \quad y = \sin\phi \sin\theta , \quad z = \cos\theta, \quad X = \frac{\sin\theta}{1-\cos\theta}\cos\phi, \quad Y = \frac{\sin\theta}{1-\cos\theta}\sin\phi
\end{align*}
On $S^2$, the angles $(\theta,\phi)$ are the latitude and longitude coordinates respectively, and $\phi$ is also the polar angle on the plane $\mathbb{R}^2$. The north pole corresponds to $\theta\rightarrow 0$. The expression for $U$ is now
\begin{align*}
\begin{pmatrix}
U_1 \\
U_2 \\
U_3
\end{pmatrix} = 
\begin{pmatrix}
\left(1-\cos\theta\right)\left(\sin^2\phi - \cos^2\phi\cos\theta\right) && -\frac{1}{2}\sin2\phi\sin^2\theta\\
-\frac{1}{2}\sin2\phi\sin^2\theta && \left(1-\cos\theta\right)\left(\cos^2\phi - \sin^2\phi\cos\theta\right)\\
\left(1-\cos\theta\right)\cos\phi \sin\theta && \left(1-\cos\theta\right)\sin\phi \sin\theta
\end{pmatrix}
\begin{pmatrix}
V_1 \\
V_2
\end{pmatrix}
\end{align*}
Suppose that the vector field $V(X,Y)$ asymptotically behaves like a power of $X-iY$. Setting $V_1+iV_2 = (X-iY)^d$ and focusing on the $\theta\rightarrow 0$ limit gives
\begin{align*}
V_1 &= \frac{\sin^d\theta}{(1-\cos\theta)^d}\cos d\phi = 2^d \theta^{-d}\cos d\phi + O(\theta^{-d+1})\\ 
V_2 &= -\frac{\sin^d\theta}{(1-\cos\theta)^d}\sin d\phi = -2^d\theta^{-d}\sin d\phi + O(\theta^{-d+1}) 
\end{align*}
To leading order in $\theta$, the vector field $U$ is
\begin{align*}
\begin{pmatrix}
U_1 \\
U_2 \\
U_3
\end{pmatrix} = \frac{1}{2}
\begin{pmatrix}
-\theta^2\cos 2\phi && -\theta^2\sin 2\phi\\
-\theta^2 \sin 2\phi && \phantom{-}\theta^2\cos 2\phi\\
\phantom{-}\theta^3\cos\phi && \phantom{-}\theta^3\sin\phi
\end{pmatrix}
\begin{pmatrix}
2^d \theta^{-d}\cos d\phi \\
-2^d\theta^{-d}\sin d\phi
\end{pmatrix} = 2^{d-1}\theta^{2-d}\left(1+O(\theta)\right)
\begin{pmatrix}
-\cos \left[(d+2)\phi\right]\\
-\sin \left[(d+2)\phi\right]\\
\theta\cos \left[(d-1)\phi\right]
\end{pmatrix}
\end{align*}
In particular, the $\theta$ dependence of $U$ as $\theta\rightarrow 0$ is 
\begin{align*}
U_1 \sim \theta^{2-d} , \qquad U_2\sim \theta^{2-d},\qquad U_3 \sim\theta^{3-d}
\end{align*}
There are therefore 3 possibilities for the behavior of $U$ at the north pole. For $d<2$, we have $|U|\rightarrow 0$, and for $d>2$, we have $|U|\rightarrow \infty$. When $d=2$, however, $U$ can be bounded but discontinuous at the north pole. 

The 3-qubit states shown in Figs.~\ref{SI_qft1} and \ref{SI_qft2} are linear combinations of planar vector fields $V$ given by
\begin{align*}
V_1 - iV_2 = h_{\boldsymbol{\sigma}}(X+iY)^3 = h_{\boldsymbol{\sigma}}(Z)^3 = \left[(Z+1)^{2\sigma_1-1}(Z-i)^{2\sigma_2 -1} (Z-1)^{2\sigma_3 -1}\right]^3 \sim Z^{6(\sigma_1+\sigma_2+\sigma_3)-9}
\end{align*}
where $Z = X+iY$. The exponent $6(\sigma_1+\sigma_2+\sigma_3)-9$ is less than $2$ when at most one of the $\sigma_i$'s is equal to $1$. Therefore, the $S^2$ vector fields corresponding to the basis states $\{|000\rangle, |001\rangle, |010\rangle, |100\rangle\}$ have zeros at the north pole. For the remaining 4 basis states, $\{|111\rangle, |110\rangle, |101\rangle, |011\rangle\}$, the $S^2$ vector fields have magnitude $|U|\rightarrow \infty$ at the north pole. The quantum Fourier transform of a computational basis state, $|\sigma_1\sigma_2\sigma_3\rangle$, produces a state which has nonzero overlap with every basis state. As a result, the $S^2$ vector fields for the states $QFT|\sigma_1\sigma_2\sigma_3\rangle$ all have magnitude $|U|\rightarrow \infty$ at the north pole (Figs.~\ref{SI_qft1} and \ref{SI_qft2}).

\bibliography{qbit_references}